\documentclass[preprint,10pt]{elsarticle}
\usepackage{amsfonts}
\usepackage{epsfig,amsmath,latexsym,amssymb}
\usepackage{amscd}
\usepackage{multirow}
\usepackage{graphicx}
\usepackage{geometry}
\usepackage{lscape}
\usepackage{amsmath}
\usepackage{amssymb}
\usepackage{bm}
\usepackage{subfigure}

\providecommand{\U}[1]{\protect\rule{.1in}{.1in}}
\newcommand{\Remm}[1]{}

\newtheorem{model ass}[theo]{Model Assumptions}

\numberwithin{equation}{section}

\begin{document}

\begin{frontmatter}

\title{Analytic Loss Distributional Approach Models for Operational Risk from the $\alpha$-Stable Doubly Stochastic Compound Processes and Implications for Capital Allocation.}
\author{Gareth W.~Peters$^{1}$ \quad Pavel .~Shevchenko$^{2}$ \quad Mark .~Young$^{3}$ \quad Wendy .~Yip$^{3}$} 
\date{{\footnotesize {Working paper, version from \today }}}
\maketitle

\begin{abstract}
\noindent Under the Basel II standards, the Operational Risk (OpRisk) advanced measurement approach is not prescriptive regarding the class of statistical model utilised to undertake capital estimation. It has however become well accepted to utlise a Loss Distributional Approach (LDA) paradigm to model the individual OpRisk loss process corresponding to the Basel II Business line/event type. In this paper we derive a novel class of doubly stochastic $\alpha$-stable family LDA models. These models provide the ability to capture the heavy tailed loss process typical of OpRisk whilst also providing analytic expressions for the compound process annual loss density and distributions as well as the aggregated compound process annual loss models. In particular we develop models of the annual loss process in two scenarios. The first scenario considers the loss process with a stochastic intensity parameter, resulting in an inhomogeneous compound Poisson processes annually. The resulting arrival process of losses under such a model will have independent counts over increments within the year. The second scenario considers discretization of the annual loss process into monthly increments with dependent time increments as captured by a Binomial process with a stochastic probability of success changing annually. Each of these models will be coupled under an LDA framework with heavy-tailed severity models comprised of $\alpha$-stable severities for the loss amounts per loss event. In this paper we will derive analytic results for the annual loss distribution density and distribution under each of these models and study their properties.
\end{abstract}


\begin{center}
{\footnotesize {\ \textit{$^{1}$ UNSW Mathematics and Statistics
Department, Sydney, 2052, Australia; \\[0pt]
email: garethpeters@unsw.edu.au \\[0pt]
(Corresponding Author) \\[0pt]
$^{2}$ CSIRO CMIS Sydney, NSW, Australia \\[0pt]
$^{3}$ Risk Services, Deloitte Touche Tohmatsu, Sydney, NSW, Australia \\[0pt] } } }
\end{center}

\end{frontmatter}

\pagebreak

\section{Motivation}
The modelling of operational risk (OpRisk) has taken a prominent place in financial quantitative measurement, this has occurred as a result of Basel II / Basel III regulatory requirements. As a result Operational risk (OpRisk) has become increasingly important to the banking industry to address these regulatory standards in light of globalization, complex financial products and changes in information technology, combined with a growing number of high-profile operational loss events worldwide. 

There was no widely accepted definition of OpRisk when the Basel Committee on Banking Supervision (BCBS) began discussions on OpRisk management at the end of the 1990s; see BCBS (1998). Often, OpRisk was defined as any risk not categorised as market or credit risk. Some banks defined it as the risk of loss arising from various types of human or technical error. Some earlier definitions can be found in a 1997 survey conducted by the British Bankers Association (BBA). In January 2001, the Basel Committee on Banking Supervision issued a proposal for a New Basel Capital Accord (referred to as Basel II) where OpRisk was formally defined as a new category of risk, in addition to market and credit risks, attracting a capital charge. In the working paper BCBS (September 2001) on the regulatory treatment of OpRisk and in the revised Basel II framework BCBS (2004), the following definition of OpRisk was adopted. \textquotedblleft \textit{Operational risk is defined as the risk of loss resulting from inadequate or failed internal processes, people and systems or from external events. This definition includes legal risk but excludes strategic and reputational risk.} \textquotedblright\ This definition did not change in the latest version of Basel II framework, BCBS (2006, p. 144). The International Actuarial Association, IAA (2004), has adopted the same definition of operational risk in the capital requirements for insurance companies. 

So OpRisk is indeed a broad category. The BCBS gives a further classification into seven event types of OpRisk (BCBS, 2006, Annex 9): Internal Fraud; External Fraud; Employment Practices and Workplace Safety; Clients, Products and Business Practices; Damage to Physical Assets; Business Disruption and System Failure; Execution, Delivery and Process Management; which serves to further illustrate the disparate nature of events in this class. Reputational and strategic risk do not fall under the OpRisk umbrella, and market and credit risks are treated separately, but almost any other event that may result in a loss to a bank, including legal action, may be termed OpRisk. 

Basel II considers three pillars, which, by their very nature, emphasize the importance of assessing, modelling and understanding OpRisk profiles. These three pillars are \textsl{minimum capital requirements} (refining and enhancing risk modelling frameworks), \textsl{supervisory review} of an institution's capital adequacy and internal assessment processes and \textsl{market discipline}, which deals with disclosure of information. Since this time, the discipline of OpRisk and its quantification have grown in prominence in the financial sector.

To illustrate just how significant OpRisk can be to a financial institution, one only needs to consider OpRisk related events such as the 1995 Barings Bank loss of around 1.3 billion GBP; the 2001 Enron loss of around 2.2 billion USD; the 2004 National Australia Bank loss of 360m AUD; and the more recent Societe Generale loss of 4.9 billion Euro. Each of which demonstrates the severity of loss process that should be modelled by OpRisk statistical models, providing strong motivation for heavy-tailed loss process models such as those involving infinite mean and variace as captured by the family of $\alpha$-stable models considered in this paper.

The impact that such significant losses have had on the financial industry and its perceived stability combined with the Basel II regulatory requirements have significantly changed the view, that financial institutions have regarding OpRisk. Under the three pillars of the Basel II agreement, set out in the framework, internationally active banks are required to set aside capital reserves against risk, to implement risk management frameworks and processes for their continual review, and to adhere to certain disclosure requirements. 

Whilst many OpRisk events occur frequently and with low impact (indeed, are `expected losses'), others are rare, and their impact may be as extreme as the total collapse of the bank. The modelling and development of methodology to capture, classify and understand properties of operational losses is a new research area in the banking and finance sector.

There are three broad approaches, that a bank may use to calculate its minimal capital reserve, as specified in the first pillar of the Basel II agreement. They are known as Basic Indicator Approach, Alternative Standardized Approach and Advanced Measurement Approach (AMA). In this paper the approach considered is the AMA. AMA is of interest since it is the most advanced framework with regards to statistical modelling.

A bank adopting the AMA must develop a comprehensive internal risk quantification system. This approach is the most flexible from a quantitative perspective, as banks may use a variety of methods and models, they believe are most suitable for their operating environment and culture, provided they can convince the local regulator, [BCBS, 2006, p150-2]. The key quantitative criteria is that a bank's models must sufficiently account for potentially high-impact rare events. In this paper we consider the idea of the loss distribution approach (LDA) which involves modelling the severity and frequency distributions over a predetermined time horizon, typically annual as specified in the APS115 section on soundness standards.

The fitting of frequency and severity distributions, as opposed to simply fitting a single parametric annual loss distribution, involves making the mathematical choice of working with compound distributions. This would seem to complicate the matter, since it is well known, that for most situations, analytical expressions for the distribution of a compound random variable are not attainable. However, the special classes of $\alpha$-Stable models developed in this paper overcome this complication.

Typically, the reason for modelling severity and frequency distributions separately then constructing a compound process because some factors affect the frequency and others may affect the severity, see\cite{Panjer06}. Some of the key points relating to why this is important in most practical settings are that the expected number of operational losses will change as the company grows. Typically growth needs to be accounted for in forecasting the number of OpRisk losses in future years, based on previous years. This can easily be understood, when modelling is performed for frequency and severity separately. Economic inflationary effects can be directly factored into size of losses through scaling of the severity distribution. Insurance and the impacts of altering policy limits and excesses are easily understood by directly altering severity distributions. Changing recording thresholds for loss events and the impact this will have on the number of losses required to be recorded is transparent.

The most popular choices for frequency distributions are Poisson, binomial and negative binomial. The typical choices of severity distribution include exponential, Weibull, lognormal, generalized Pareto, the $g$-and-$h$ family of distributions \cite{Dutta06}, \cite{Peters06} and recently the $\alpha$-Stable family \cite{Peters10}.

The most important processes to model accurately are those, which have relatively infrequent losses. However, when these losses do occur they are distributed as a very heavy-tailed severity distribution. In particular we focus our analysis on the scenarios involving heavy-tailed severity models in the rare-event extreme consequence context. Thereby providing analysis of the loss process most likely to have significant consequences on a financial institution, those which may lead to ruin. This involves introducing to OpRisk modeling an important family of severity models, utilized in insurance claims reserving in \cite{adler1998practical}, given by the $\alpha$-stable severity model. This family of severity model are flexible enough to incorporate light-tailed Gaussian loss models through to infinite mean, infinite variance severity loss models such as the Cauchy model. 

There are many approaches, which can be used to fit and incorporate expert opinion / scenario analysis for these parametric distributions and the approach adopted by a bank will depend on the data source being modelled and how much confidence one has in the data source. After the best-fitting models are selected, these are combined to produce a compound process for the annual loss distribution. From this compound process, VaR and capital estimates may be derived.

Once compound processes have been fitted for each business unit and risk type, the next step in the process is to aggregate these annual loss random variables for each individual {\{}business line-event type{\}} combination, and thus to obtain the institution-wide annual loss distribution. This paper will not address the issues associated with correlation and dependence modelling. For more information on typical approaches to introducing correlation in an aggregation process, including copula methods, correlation of frequency, severity or annual losses, see \cite{Cruz04} and \cite{peters2009dynamic}. In the next section we present the details of the LDA modelling framework adopted in this paper.

\section{Loss Distributional Approach Model Specifications}
\label{LDA_Models}
OpRisk LDA models are discussed widely in the literature; see e.g. \cite{Cruz04}, \cite{Chavez-Demoulin} \cite{Frachot}, \cite{Shevchenko09}. Under the LDA Basel II requirements, banks should quantify distributions for frequency and severity of OpRisk for each business line and event type over a one-year time horizon. In this section we begin by presenting a generic LDA framework before presenting specific models that we will develop under the compound process $\alpha$-Stable family. The standard LDA Basel II structures, involving an annual loss in a risk cell (business line/event type) modeled as a compound random variable,%
\begin{equation}
Z_{t}^{\left( j\right) }=\sum\limits_{s=1}^{N_{t}^{\left( j\right)
}}X_{s}^{\left( j\right) }\left( t\right).  
\label{AnnLoss1}
\end{equation}%
Here $t=1,2,\ldots,T,T+1$ in our framework is discrete time (in annual units) with $T+1$ corresponding to the next year. The upper script $j$ is used to identify the risk cell. The annual number of events $N_{t}^{(j)}$ is a random variable distributed according to a frequency counting distribution $P^{(j)}(\cdot) $, typically Poisson. The severities in year $t$ are represented by random variables $X_{s}^{(j)}(t)$, $s\ge1$, distributed according to a severity distribution $F^{(j)}(\cdot)$, typically lognormal. Severities represent actual loss amounts per event. 

To reflect both the nature of OpRisk data with extreme but rare events, the severity models selected for this analysis are chosen to exhibit extreme heavy-tails, with particular interest in distributions with infinite mean/variance. To accomplish this we consider the family of $\alpha$-Stable severity distributions. These particular models have been proposed as suitable models for insurance claims modeling and finance in previous papers, such as \cite{embrechts1994modeling}, \cite{mcneil2000estimation}, \cite{fama1968some}, \cite{peters2010bayesian} and \cite{peters2009likelihood}.

The study of the distribution of the annual loss process is one of the classical problems in risk theory. Closed-form solutions are typically not available for the distributions used in OpRisk. However with modern computer processing power, these distributions can be calculated virtually exactly using numerical algorithms, see approaches in \cite{peters2007simulation}. The easiest to implement is the Monte Carlo method. However, because it is typically slow, Panjer recursion and Fourier inversion techniques are widely used. Both have a long history, but their applications to computing very high quantiles of the compound distribution functions with high frequencies and heavy tails are only recent developments and various pitfalls exist. In this paper the class of OpRisk models developed will be proven to admit closed form analytic solutions to the density and distribution functions of the annual loss and of the total bank's loss, for independent risks, in year $t$ which is calculated as
\begin{equation}
Z_{t}=\sum\limits_{j=1}^{J}Z_{t}^{\left( j\right) },
\label{totalLoss}
\end{equation}%
where formally for OpRisk under the Basel II requirements $J=56$ (seven event types times eight business lines). However, this may differ depending on the financial institution and type of problem. In this paper we will drop the index $j$ for the risk cell unless required.

The approach we consider in this paper utilizes the representation of the annual loss process for a given business line and event type combination given by a convolution. In particular, it is well known that the density and distribution of the sum of two independent continous random variables $Y_1 \sim F_1(\cdot)$ and $Y_2 \sim F_2(\cdot)$ with densities $f_1(\cdot)$ and $f_2(\cdot)$ respectively, can be calculated via convolution as
$$f_{Y_1 + Y_2}(y) = \left(f_1 \star f_2\right)(y) = \int f_2(y-y_1)f_1(y_1)dy_1$$
and
$$F_{Y_1 + Y_2}(y) = \left(F_1 \star F_2\right)(y) = \int F_2(y-y_1)f_1(y_1)dy_1$$
respectively. Here the notation $f_1 \star f_2$ denotes the convolution of $f_1$ and $f_2$ functions as above and the notation $Y \sim F(y)$ means a random variable has a distribution function $F(y)$. Thus the distribution of the aggregated loss in Equation \ref{AnnLoss1} can be calculated via convolutions as
\begin{equation}
\begin{split}
F(z) &= Pr\left[Z \leq z\right] = \sum_{n=0}^{\infty} Pr\left[Z \leq z| N=n\right] Pr\left[N=n\right] \\
&= \sum_{n=0}^{\infty} p_n F^{(n)\star}(z).
\end{split}
\end{equation}
Here, $F^{(n)\star}(z) = Pr\left[X_1 + X_2 + \cdots + X_n \leq z\right]$ is the $n$-th convolution of $F(\cdot)$ calculated recursively as
$$F^{(n)\star}(z) = \int_0^z F^{(n-1)\star}(z)(z-x) f(x) dx$$
with $F^{(0)\star}(z) = 1$ if $z \geq 0$ and zero otherwise.

Next, we present a special family of statistical models for the severity distribution which are well known to be closed under convolution, meaning the analytic solution to the recursive convolution integrals above are known in closed analytic parameteric form for both the distribution and densities.

\subsection{$\alpha$-Stable Severity Distribution Models}
Considered as generalizations of the Gaussian distribution, $\alpha$-Stable models are defined as the class of location-scale distributions which are closed under convolutions. As in \cite{Peters10} we restrict to the class of truncated $\alpha$-stable models to ensure we only work with non-negative loss processes. In an OpRisk context, $\alpha$-stable distributions possess several useful properties, including infinite mean and infinite variance, skewness and heavy tails \cite{zolotarev1986one} and \cite{samorodnitsky1994stable}. 

We assume the $i$-th loss of the $j$-th risk process in year $t$ in a risk cell is a random variable with $\alpha$-stable distribution, denoted by $X^{(j)}_i(t) \sim S_{\alpha}\left(x; \beta, \gamma, \delta, 0\right)$. Where, $S_{\alpha}\left(x; \beta, \gamma, \delta, 0\right)$ denotes the univariate four parameter stable distribution family under parameterization $S(0)$, see algorithm in the Appendix and details contained in \cite{peters2009likelihood}.  

The univariate $\alpha$-stable distribution we consider is specified by four parameters: $\alpha \in (0, 2]$ determining the rate of tail decay; $\beta \in [-1, 1]$ determining the degree and sign of asymmetry (skewness); $\gamma > 0$ the scale (under some parameterizations); and $\delta \in \mathbb{R}$ the location. The parameter $\alpha$ is termed the characteristic exponent, with small and large $\alpha$ implying heavy and light tails respectively. Gaussian $(\alpha = 2, \beta = 0)$, Cauchy $(\alpha = 1, \beta = 0)$ and Levy $(\alpha = 0.5, \beta = 1)$ distributions provide the only analytically tractable sub-members of this family. Except these special cases, in general $\alpha$-stable models admit no closed-form expression for the density which can be evaluated point-wise, inference typically proceeds via the characteristic function, see discussions in \cite{peters2009likelihood}. However, intractable to evaluate point-wise, importantly for OpRisk applications, simulation of random variates is very efficient, see \cite{chambers1976method} and the algorithm provided in Appendix \ref{Append1}. 

From \cite{Nolan}, a random variable $X$ is said to have a stable distribution, $S(\alpha,\beta,\gamma,\delta;0)$, if its characteristic function has the following form:
\[ E[\text{exp}(i\theta X)] = 	\left\{              
			\begin{array}{ll}
				\text{exp}\{ -\gamma^\alpha|\theta|^\alpha(1+i\beta(\text{sign}(\theta))\tan({\pi \alpha \over 2})(|\gamma \theta|^{1-\alpha}-1))+i\delta \theta\} & \text{if   } \alpha \neq 1\\ 
				\text{exp}\{ -\gamma|\theta|(1+i\beta({2 \over \pi})(\text{sign}(\theta))\text{ln}(\gamma|\theta|))+i\delta \theta\} & \text{if   } \alpha = 1.
				
			\end{array}
			\right.
	\]

In the following Lemmas we present some fundamental basic facts about $\alpha$-Stable random variables that will be required to establish the novel analytic closed form expressions for the annual loss LDA models we develop in this paper. These will be used to construct analytic exact binomial, negative binomial, Poisson and doubly stochastic Poisson mixture representations of the annual loss process for a bank under $\alpha$-stable severity models with the required positive support. This will be achieved by considering a special sub-family of $\alpha$-stable models. 

In addition we will provide an analytic expression for the tail distribution of these models and the properties of the ES in special cases of this LDA model. Finally we will utilise known asymptotic results to comment on the institution wide capital aggregation results as a result of models containing dominating risk processes such as those contained in the $\alpha$-Stable family. These results will extend those developed for OpRisk insurance models recently in \cite{Peters10}.

\textbf{Lemma 1} \textit{If $Y \sim S(\alpha,\beta,\gamma,\delta;0)$, then for any $a \neq 0, b \in \Re$, the transformation $Z=aY+b$ is a scaled version of the $\alpha$-stable distribution. That is $Z \sim S(\alpha,(\text{sign}(a)\beta,|a|\gamma,a\delta+b;0)$. In addition, the characteristic functions, densities and distribution functions are jointly continuous in all four parameters $(\alpha,\beta,\gamma,\delta)$ and in $x$. These results follow from \cite{samorodnitsky1994stable} and \cite{Nolan} Proposition 1.16.}

\textbf{Lemma 2} \textit{If for all $i \in \left\{1,\ldots,N\right\}$ one has random variables $X_i \sim S(\alpha,\beta_i,\gamma_i,\delta_i;0)$ then the distribution of the linear combination, given N, is 
\begin{equation}
\begin{split}
Z &= \sum_{i=1}^N X_i \sim S(\alpha,\widetilde{\beta},\widetilde{\gamma},\widetilde{\delta};0)\\
\widetilde{\gamma}^{\alpha} &= \sum_{i=1}^N \gamma_i^{\alpha}, \; \; \; \; \widetilde{\beta} =  \frac{\sum_{i=1}^N \beta_i\gamma_i^{\alpha}}{\sum_{i=1}^N \gamma_i^{\alpha}} \; \; \; \;
\widetilde{\delta} = 	\left\{              
			\begin{array}{ll}
			\sum_{i=1}^N \delta_i + \tan \frac{\pi \alpha}{2}\left(\widetilde{\beta}\widetilde{\gamma} - \sum_{i=1}^N \beta_j\gamma_j\right) & \text{if   } \alpha \neq 1\\  
			\sum_{i=1}^N \delta_i + \frac{2}{\pi}\left(\widetilde{\beta}\widetilde{\gamma}\log\widetilde{\gamma} - \sum_{i=1}^N \beta_j\gamma_j\log\gamma_i\right) & \text{if   } \alpha = 1
			\end{array}
			\right. 
\end{split}
\end{equation}
This result follows from (\cite{samorodnitsky1994stable} Section 1.2, Property 1.2.1) and (\cite{Nolan}, Proposition 1.17).}

\begin{figure}[!ht]
\includegraphics[width = 0.8\textwidth, height = 6cm]{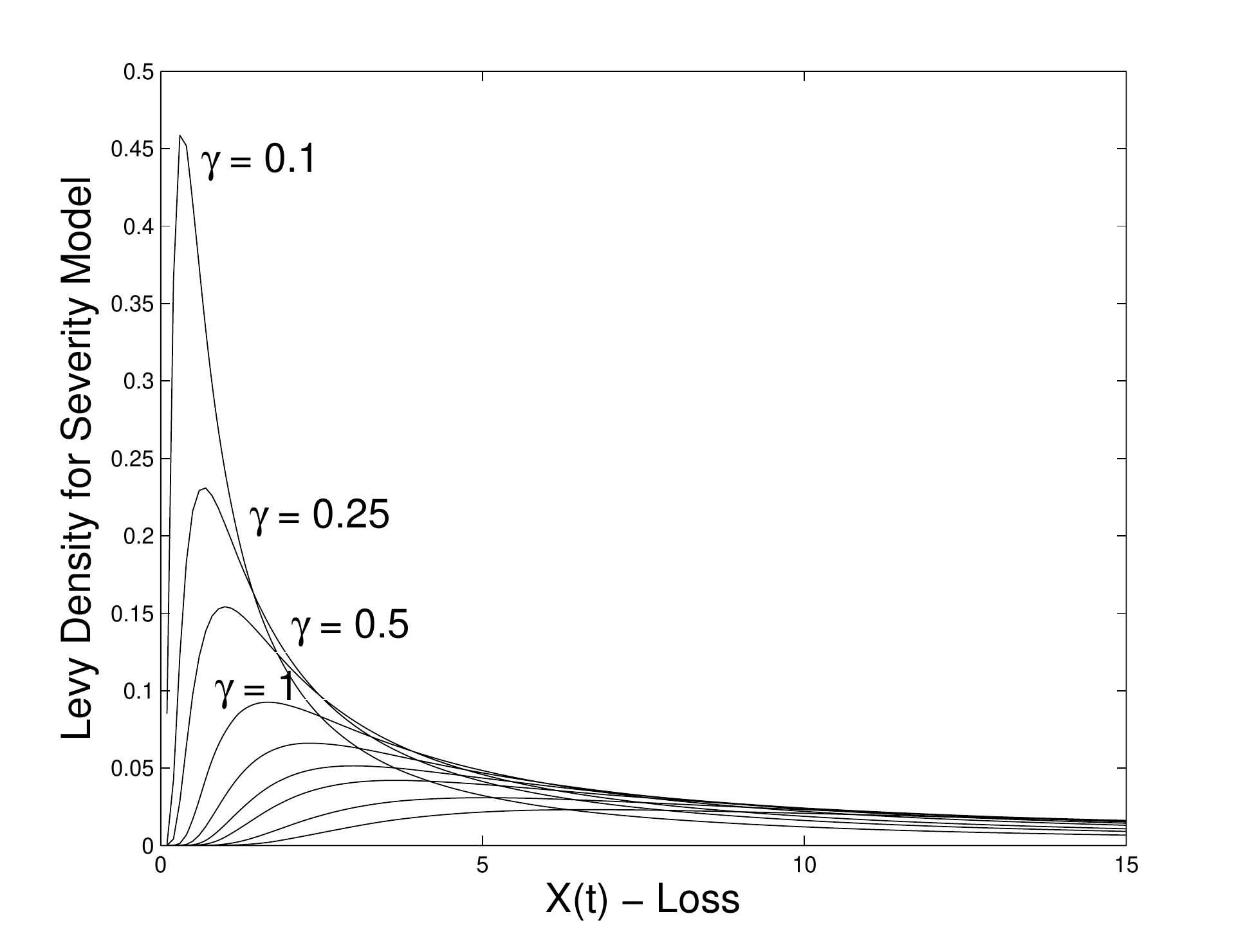}
\caption{Example of the Levy severity model as a function of scale parameter, with location $\delta = 0$.}
\label{Fig_Levy}
\end{figure}

The property in Lemma 2, of closure under convolution for random variables with identical $\alpha$ parameter is lost under truncation to positive support. The implications of this are that in OpRisk the convolution property will in general be lost for general members of the truncated severity models ($X > 0$). This is with one notable exception given by the sub-family of Levy distributions. Examples of loss distributions for this sub-family are illustrated in Figure \ref{Fig_Levy} as a function of the scale parameter $\gamma$. 
 
\textbf{Lemma 3} \textit{If $X \sim S(0.5,1,\gamma,\delta;0)$ this model specifies the sub-family of $\alpha$-stable models with positive real support $x \in [\delta,\infty)$. The density and distribution functions are analytic and given respectively, for $\delta < x < \infty$, by
\begin{equation*}
f_{X}(x) = \sqrt{\frac{\gamma}{2 \pi}}\frac{1}{\left(x-\delta\right)^{3/2}}\exp\left(-\frac{\gamma}{2\left(x-\delta\right)}\right), \; \; F_{X}(x) = \text{erfc}\left(\sqrt{\frac{\gamma}{2\left(x-\delta\right)}}\right).
\end{equation*}
The median is given by $\widetilde{\mu} = \frac{\widetilde{\gamma}_n}{2}\left(\text{erfc}^{-1}\left(0.5\right)\right)^2$ and the mode $M = \frac{\widetilde{\gamma}_n}{3}$,
where $\text{erfc}(x) = 1-\text{erf}(x) = 1-\frac{2}{\sqrt{\pi}}\int_{0}^xe^{-t^2}dt$. This result follows from \cite{Nolan} (Chapter 1. p.5).}

\subsubsection{Estimation of $\alpha$-stable severity models in OpRisk}
From the perspective of application of such $\alpha$-stable models in OpRisk LDA settings, there are efficient parameter estimation routines available given OpRisk data; see a discussion on Bayesian approaches and review in \cite{peters2009likelihood}. There are also available numerical maximum likelihood based approaches, see \cite{samorodnitsky1994stable} and the software package implementing such numerical procedures as discussed in \cite{Nolan}. In this section we discuss the practical perspective of utilising a Levy stable model for the severity distribution in an OpRisk context and how this would be justified in practical data analysis. Firstly, in an OpRisk LDA setting the severity distribution will only take positive support, hence the restriction of $\beta=1$ will provide this result by ensuring a positive skew that results in support only on the positive real line. We now discuss the appropriateness in practice of considering the Levy class of $\alpha$-stable models, which satisfy the setting in which $\beta = 1$ and importantly admit closed form expressions for the density and distribution functions to be which we demonstrate in this paper results in closed form expressions for the annual loss distribution.

We illustrate this reasoning on the following example, where we consider $N=100$ losses $X_{1:N}$ obtained from a $\alpha$-stable model $S_{\alpha=0.7}\left(1,\gamma = 0.01, \delta = 0\right)$. We fit general $\alpha$-stable models for an increasing number of annual losses $n \in \left\{5,10,15,100\right\}$, the estimated MLE parameters and associated 95\% C.I. are provided in Table \ref{Estimates}. These results were obtained utilising the software avialable on John Nolans $\alpha$-stable website \vspace{-0.7cm} \begin{verbatim}http://academic2.american.edu/~jpnolan/stable/stable.html \end{verbatim}.

\begin{table}[!h]
\begin{tabular}{|c|c|c|c|c|}\hline
MLE parameter & N = 5 & N = 10 & N = 15 & N = 100\\ 
estimates ($95\% C.I [\cdot]$) & & & &\\ \hline
$\alpha$ 	& 1.01 [0,2]				& 0.41 [0,2]					& 0.45 [0,2]					& 0.60 [0.46,0.74]\\
$\beta$ 	& 0.04 [-1,1]				& 0.99 [-1,1]					& 1.00 [-1,1]					& 1.00 [1.0,1.0] \\
$\gamma$ 	& 0.51 [-0.14,1.16] & 0.14 [-0.23, 1.02]	& 0.12 [-0.18, 0.84]	& 0.01 [0.005,0.015]  \\
$\delta$ 	& -0.34 [0.31,0.99]	& 0.12 [-0.26, 0.83]	& 0.13 [-0.12, 0.56]	& 0.02 [0.016, 0.024]\\ \hline
\end{tabular}
\label{Estimates}
\caption{MLE parameter estimates of $S_{\alpha}\left(\beta,\gamma,\delta\right)$ and 95\% C.I.}
\end{table}

The largest number of observations considered in this example with $N=100$ is already practically a very large number of observations in the context of OpRisk in which heavy-tailed severity models are suitable and we see in this example that even with 100 observed losses, fitting the heavy tailed $\alpha$-stable model estimates accurately the parameters $\beta$, $\gamma$ and $\delta$. However, estimation of the tail index, in this class of model represented by $\alpha$, with precision is a well known challenging problem with many classes of heavy tailed severity model. We demonstrate this here, where the tail index 95\% C.I remains wide even with $N=100$ observations. Therefore, when working with heavy-taile models in the $\alpha$-stable family for an OpRisk setting it is sensible to consider selecting the sub-class of $\alpha$-stable models which admit a parameteric form, this occurs for the Levy sub-family of $\alpha$-stable models as presented in Lemma 3. We will now consider the LDA framework under this sub-class of severity model and demonstrate in such classes we can obtain closed form expressions for the annual loss distribution in several settings.

Next we develop novel LDA models utilising the sub-family of $\alpha$-Stable Levy family severity models. We begin by considering discrete time loss processes in which the total number of lossess is assessed in fixed intervals of time, such as quarterly or monthly time increments. In such cases we may wish to consider non-independent counts per month and so can consider the annual loss distribution modelled by a frenquency distribution with binomial distribution or with negative binomial distribution, depending on whether we believe the variance in the number of losses exceeds the mean number of losses or is less than the mean number of losses. In the continous arrival process setting, we consider annual losses to arrive accroding to independent increments of time according to an exponential arrival process, resulting in a Poisson frequency distribution. Clearly under a Poisson based model we are implicitly assuming the mean number of losses per year is equal to the variance in the number of annual losses, when the rate parameter represents the average number of annual losses. We will then extend this Poisson family of models to the doubly stochastic family of frequency distributions and derive closed form expressions for the annual loss distribution in such cases.

\subsection{binomial and negative binomial $\alpha$-Stable LDA Analytic Models}
In this section we consider the binomial and negative binomial models where we assume that the random variable for the total number of events per year are first given by standard binomial and negative binomial Processes, before extending these results to doubly stochastic processes. For the binomial proces we obtain the following result in Theorem 1.

\textbf{Theorem 1} \textit{The distribution of the annual loss process $Z$ represented by a compound process model with LDA structure in which the frequency is $N(t) \sim Bi(M,p)$ and the severity model \\$X_i(t) \sim S(0.5,1,\gamma,\delta;0)$, then the exact density of the annual loss process can be expressed analytically as a mixture density comprised of $\alpha$-stable components with binomial mixing weights for $N_t > 0$,
\begin{equation}
f_Z(z) = \sum_{n=1}^{M} C^M_n \left(p\right)^n \left(1-p\right)^{M-n} \left[ \sqrt{\frac{\widetilde{\gamma}_n}{2\pi}}\frac{1}{\left(z-\widetilde{\delta}_n\right)^{3/2}}\exp\left(-\frac{\widetilde{\gamma}_n}{2\left(z-\widetilde{\delta}_n\right)}\right) \right] \times \mathbb{I}\left[\widetilde{\delta}_n<z<\infty\right]
\label{BinoMix}
\end{equation}
with
\begin{equation*}
\begin{split}
\widetilde{\gamma}_n^{0.5} = \sum_{i=1}^n \gamma_i^{0.5} = n|\gamma|^{0.5}, \; \; \; \; \widetilde{\beta}_n = 1 \; \; \; \;
\widetilde{\delta_n} = \sum_{i=1}^n \delta_i + \tan \frac{\pi}{4}\left(\widetilde{\gamma}_n - \sum_{j=1}^n \gamma_j\right) = n\delta + \tan \frac{\pi}{4}\left(n^2|\gamma| - n\gamma\right),  
\end{split}
\end{equation*}
and $F_Z(0) = \text{Pr}(N_t=0)=\left(1-p\right)^{M}$ for $N=0$. The exact form of the annual loss cumulative distribution function is also expressible in closed-form,
\begin{equation}
\begin{split}
\Pr\left(Z < z\right) = F_Z(z) &= \sum_{n=1}^{M} C^M_n \left(p\right)^n \left(1-p\right)^{M-n} \text{erfc}\left( \sqrt{\frac{\widetilde{\gamma}_n}{2\left(z-\widetilde{\delta}_n\right)}}\right)\times \mathbb{I}\left[\widetilde{\delta}_n<z<\infty\right] \\
&+ \left(1-p\right)^{M}. 
\label{Bino2Mix}
\end{split}
\end{equation}
This result follows directly from application of Lemma 1, Lemma 2 and Lemma 3.}

For the doubly stochastic binomial-Beta proces we can state the following result in Theorem 2.

\textbf{Theorem 2} \textit{The distribution of the annual loss process $Z$ represented by a compound process model with LDA structure in which the frequency is $N(t) \sim Bi\left(M,p\right)$, $p \sim Be\left(\alpha,\beta\right)$  and the severity model $X_i(t) \sim S(0.5,1,\gamma,\delta;0)$, then the exact density of the annual loss process can be expressed analytically as a mixture density comprised of $\alpha$-stable components with Beta mixing weights for $N_t > 0$,
\begin{equation}
\begin{split}
f_Z(z) &= \sum_{n=1}^{M} \int_0^1 C^M_n \left(p\right)^n \left(1-p\right)^{M-n} \frac{1}{B\left(\alpha,\beta\right)} p^{\alpha-1}(1-p)^{\beta-1} dp \\ 
&\left[ \sqrt{\frac{\widetilde{\gamma}_n}{2\pi}}\frac{1}{\left(z-\widetilde{\delta}_n\right)^{3/2}}\exp\left(-\frac{\widetilde{\gamma}_n}{2\left(z-\widetilde{\delta}_n\right)}\right) \right] \times \mathbb{I}\left[\widetilde{\delta}_n<z<\infty\right]\\
&= \sum_{n=1}^{M} \frac{M!B(\alpha + n, \beta + M - n)}{(M-n)!n!B(\alpha,\beta)} \left[ \sqrt{\frac{\widetilde{\gamma}_n}{2\pi}}\frac{1}{\left(z-\widetilde{\delta}_n\right)^{3/2}}\exp\left(-\frac{\widetilde{\gamma}_n}{2\left(z-\widetilde{\delta}_n\right)}\right) \right] \times \mathbb{I}\left[\widetilde{\delta}_n<z<\infty\right]
\end{split}
\label{BinoBetaMix}
\end{equation}
with
\begin{equation*}
\begin{split}
\widetilde{\gamma_n}^{0.5} = \sum_{i=1}^n \gamma_i^{0.5} = n|\gamma|^{0.5}, \; \; \; \; \widetilde{\beta_n} = 1 \; \; \; \;
\widetilde{\delta_n} = \sum_{i=1}^n \delta_i + \tan \frac{\pi}{4}\left(\widetilde{\gamma}_n - \sum_{j=1}^n \gamma_j\right) = n\delta + \tan \frac{\pi}{4}\left(n^2|\gamma| - n\gamma\right),  
\end{split}
\end{equation*}
and $F_Z(0) = \text{Pr}(N_t=0)= \frac{B(\alpha, \beta + M)}{B(\alpha,\beta)}$ for $N=0$. Note we also denote the standard Beta function by $B(x,y) = \int_0^1 t^{x-1}(1-t)^{y-1}dt$. The exact form of the annual loss cumulative distribution function is also expressible in closed-form,
\begin{equation}
\begin{split}
\Pr\left(Z < z\right) = F_Z(z) &= \sum_{n=1}^{M} \frac{M!B(\alpha + n, \beta + M - n)}{(M-n)!n!B(\alpha,\beta)} \text{erfc}\left( \sqrt{\frac{\widetilde{\gamma}_n}{2\left(z-\widetilde{\delta}_n\right)}}\right)\times \mathbb{I}\left[\widetilde{\delta}_n<z<\infty\right] \\
&+ \frac{B(\alpha, \beta + M)}{B(\alpha,\beta)}. 
\label{BinoBeta2Mix}
\end{split}
\end{equation}}
\textbf{Proof} \hspace{0.5cm} The result for the convolution between the stable random variables, conditional on $n$, follows from application of Lemma 1, Lemma 2 and Lemma 3. The mixing weight of the compound process for the binomial-Beta distribution is derived by considering Bayes Theorem and the conjugacy property of the binomial-Beta model from which we know $p(p|n) = Be(p; \alpha + n, \beta + M - n)$, hence
\begin{equation}
\begin{split}
\mathbb{P}\text{r}(n) &= \frac{p(n|M,p)p(p)}{p(p|n)} =\frac{Bi(n;M,p)Be(p;\alpha,\beta)}{Be(p; \alpha + n, \beta + M - n)} = \frac{\frac{M!}{(M-n)!n!}p^n(1-p)^{(M-n)}\frac{p^{(\alpha-1)}(1-p)^{(\beta-1)}}{B(\alpha,\beta)}}{\frac{p^{(\alpha + n - 1)}(1-p)^{(\beta + M -n -1)}}{B(\alpha + n,\beta+M-n)}}  \\
&= \frac{M!B(\alpha + n, \beta + M - n)}{(M-n)!n!B(\alpha,\beta)}. \text{\hspace{2cm} \qed}
\end{split}
\end{equation}

For the negative binomial proces we obtain the following result in Theorem 3.

\textbf{Theorem 3} \textit{The distribution of the annual loss process $Z$ represented by a compound process model with LDA structure in which the frequency is $N(t) \sim NB(r,p)$ and the severity model \\$X_i(t) \sim S(0.5,1,\gamma,\delta;0)$, then the exact density of the annual loss process can be expressed analytically as a mixture density comprised of $\alpha$-stable components with negative binomial mixing weights for $N_t > 0$,
\begin{equation}
f_Z(z) = \sum_{n=1}^{\infty} C^{n+r-1}_{n} \left(1-p\right)^(n) \left(p\right)^{r} \left[ \sqrt{\frac{\widetilde{\gamma}_n}{2\pi}}\frac{1}{\left(z-\widetilde{\delta}_n\right)^{3/2}}\exp\left(-\frac{\widetilde{\gamma}_n}{2\left(z-\widetilde{\delta}_n\right)}\right) \right] \times \mathbb{I}\left[\widetilde{\delta}_n<z<\infty\right]
\label{NBinoMix}
\end{equation}
with
\begin{equation*}
\begin{split}
\widetilde{\gamma_n}^{0.5} = \sum_{i=1}^n \gamma_i^{0.5} = n|\gamma|^{0.5}, \; \; \; \; \widetilde{\beta_n} = 1 \; \; \; \;
\widetilde{\delta_n} = \sum_{i=1}^n \delta_i + \tan \frac{\pi}{4}\left(\widetilde{\gamma}_n - \sum_{j=1}^n \gamma_j\right) = n\delta + \tan \frac{\pi}{4}\left(n^2|\gamma| - n\gamma\right),  
\end{split}
\end{equation*}
and $F_Z(0) = \text{Pr}(N_t=0)=\left(1-p\right)^{n}$ for $N=0$. The exact form of the annual loss cumulative distribution function is also expressible in closed-form,
\begin{equation}
\begin{split}
\Pr\left(Z < z\right) = F_Z(z) &= \sum_{n=1}^{\infty} C^{n+r-1}_{n} \left(1-p\right)^r \left(p\right)^{n} \text{erfc}\left( \sqrt{\frac{\widetilde{\gamma}_n}{2\left(z-\widetilde{\delta}_n\right)}}\right)\times \mathbb{I}\left[\widetilde{\delta}_n<z<\infty\right] \\
&+ \left(1-p\right)^{r}. 
\label{NBino2Mix}
\end{split}
\end{equation}
}

For the doubly stochastic negative binomial-Beta proces we can state the following result in Theorem 4.

\textbf{Theorem 4} \textit{The distribution of the annual loss process $Z$ represented by a compound process model with LDA structure in which the frequency is $N(t) \sim NB\left(r,p\right)$, $p \sim Be\left(\alpha,\beta\right)$  and the severity model $X_i(t) \sim S(0.5,1,\gamma,\delta;0)$, then the exact density of the annual loss process can be expressed analytically as a mixture density comprised of $\alpha$-stable components with Beta mixing weights for $N_t > 0$,
{\small{
\begin{equation}
\begin{split}
f_Z(z) &= \sum_{n=1}^{\infty} \int_0^1 C^{n+r-1}_{n} \left(1-p\right)^r \left(p\right)^{n} \frac{1}{B\left(\alpha,\beta\right)} p^{\alpha-1}(1-p)^{\beta-1} dp\\ &\left[ \sqrt{\frac{\widetilde{\gamma}_n}{2\pi}}\frac{1}{\left(z-\widetilde{\delta}_n\right)^{3/2}}\exp\left(-\frac{\widetilde{\gamma}_n}{2\left(z-\widetilde{\delta}_n\right)}\right) \right] \times \mathbb{I}\left[\widetilde{\delta}_n<z<\infty\right]\\
&= \sum_{n=1}^{\infty}\frac{(n+r-1)!p^{(\alpha + r -1)}B(\alpha + rn, \beta + n)}{(r-1)!n!p^{(\alpha + rn -1)}B(\alpha,\beta)} \left[ \sqrt{\frac{\widetilde{\gamma}_n}{2\pi}}\frac{1}{\left(z-\widetilde{\delta}_n\right)^{3/2}}\exp\left(-\frac{\widetilde{\gamma}_n}{2\left(z-\widetilde{\delta}_n\right)}\right) \right] \times \mathbb{I}\left[\widetilde{\delta}_n<z<\infty\right]
\label{NBinoBetaMix}
\end{split}
\end{equation}
}}
with
\begin{equation*}
\begin{split}
\widetilde{\gamma_n}^{0.5} = \sum_{i=1}^n \gamma_i^{0.5} = n|\gamma|^{0.5}, \; \; \; \; \widetilde{\beta_n} = 1 \; \; \; \;
\widetilde{\delta_n} = \sum_{i=1}^n \delta_i + \tan \frac{\pi}{4}\left(\widetilde{\gamma}_n - \sum_{j=1}^n \gamma_j\right) = n\delta + \tan \frac{\pi}{4}\left(n^2|\gamma| - n\gamma\right),  
\end{split}
\end{equation*}
and $F_Z(0) = \text{Pr}(N_t=0)=\frac{p^{(\alpha + r -1)}}{p^{(\alpha -1)}}$ for $N=0$. The exact form of the annual loss cumulative distribution function is also expressible in closed-form,
{\small{
\begin{equation}
\begin{split}
\Pr\left(Z < z\right) = F_Z(z) &= \sum_{n=1}^{\infty} \frac{(n+r-1)!p^{(\alpha + r -1)}B(\alpha + rn, \beta + n)}{(r-1)!n!p^{(\alpha + rn -1)}B(\alpha,\beta)} \text{erfc}\left( \sqrt{\frac{\widetilde{\gamma}_n}{2\left(z-\widetilde{\delta}_n\right)}}\right)\times \mathbb{I}\left[\widetilde{\delta}_n<z<\infty\right] \\
&+ \frac{p^{(\alpha + r -1)}}{p^{(\alpha -1)}}. 
\label{NBinoBetaMix2}
\end{split}
\end{equation}}
}}
\textbf{Proof} \hspace{0.5cm} The result for the convolution between the stable random variables, conditional on $n$, follows from application of Lemma 1, Lemma 2 and Lemma 3. The mixing weight of the compound process for the negative binomial-Beta distribution is derived by considering Bayes Theorem and the conjugacy property of the binomial-Beta model from which we know $p(p|n) = Be(p; \alpha + rn, \beta + n)$, hence
\begin{equation}
\begin{split}
\mathbb{P}\text{r}(n) &= \frac{p(n|r,p)p(p)}{p(p|n)} =\frac{NB(n;r,p)Be(p;\alpha,\beta)}{Be(p; \alpha + rn, \beta + n)} = \frac{C^{n+r-1}_{n} \left(1-p\right)^n\left(p\right)^{r}\frac{p^{(\alpha - 1)}(1-p)^{(\beta - 1)}}{B(\alpha,\beta)}}{\frac{p^{(\alpha + rn - 1)}(1-p)^{(\beta + n -1)}}{B(\alpha + rn,\beta+n)}}  \\
&= \frac{(n+r-1)!p^{(\alpha + r -1)}B(\alpha + rn, \beta + n)}{(r-1)!n!p^{(\alpha + rn -1)}B(\alpha,\beta)}.  \text{\hspace{2cm} \qed}
\end{split}
\end{equation}

\subsection{Poisson and Doubly Stochastic Poisson-Gamma - $\alpha$-Stable LDA Analytic Models}
In this section we begin by reviewing some results recently developed in \cite{Peters10} before extending these results to the class of doubly stochastic Poisson-Gamma-$\alpha$-Stable LDA models. Hence, we begin with the standard Poisson process LDA model result given in Theorem 5.

\textbf{Theorem 5} \textit{The distribution of the annual loss process $Z$ represented by a compound process model with LDA structure in which the frequency is $N(t) \sim Po(\lambda)$ and the severity model \\$X_i(t) \sim S(0.5,1,\gamma,\delta;0)$, then the exact density of the annual loss process can be expressed analytically as a mixture density comprised of $\alpha$-stable components with Poisson mixing weights for $N_t > 0$,
\begin{equation}
f_Z(z) = \sum_{n=1}^{\infty} \exp(-\lambda)\frac{\lambda^n}{n!} \left[ \sqrt{\frac{\widetilde{\gamma}_n}{2\pi}}\frac{1}{\left(z-\widetilde{\delta}_n\right)^{3/2}}\exp\left(-\frac{\widetilde{\gamma}_n}{2\left(z-\widetilde{\delta}_n\right)}\right) \right] \times \mathbb{I}\left[\widetilde{\delta}_n<z<\infty\right]
\label{PoissMix}
\end{equation}
with
\begin{equation*}
\begin{split}
\widetilde{\gamma_n}^{0.5} = \sum_{i=1}^n \gamma_i^{0.5} = n|\gamma|^{0.5}, \; \; \; \; \widetilde{\beta_n} = 1 \; \; \; \;
\widetilde{\delta_n} = \sum_{i=1}^n \delta_i + \tan \frac{\pi}{4}\left(\widetilde{\gamma}_n - \sum_{j=1}^n \gamma_j\right) = n\delta + \tan \frac{\pi}{4}\left(n^2|\gamma| - n\gamma\right),  
\end{split}
\end{equation*}
and $F_Z(0) = \text{Pr}(N_t=0)=\exp(-\lambda)$ for $N=0$. The exact form of the annual loss cumulative distribution function is also expressible in closed-form,
\begin{equation}
\Pr\left(Z < z\right) = F_Z(z) = \sum_{n=1}^{\infty} \exp(-\lambda)\frac{\lambda^n}{n!} \text{erfc}\left( \sqrt{\frac{\widetilde{\gamma}_n}{2\left(z-\widetilde{\delta}_n\right)}}\right)\times \mathbb{I}\left[\widetilde{\delta}_n<z<\infty\right] + \exp(-\lambda). 
\label{Poiss2Mix}
\end{equation}
This result follows directly from application of Lemma 1, Lemma 2 and Lemma 3.}

We can now present results for the doubly stochastic Poisson-Gamma process LDA model, given in Theorem 6.

\textbf{Theorem 6} \textit{The distribution of the annual loss process $Z$ represented by a doubly stochastic compound process model with LDA structure in which the frequency is $N(t) \sim Po(\lambda)$ and $\lambda \sim \Gamma\left(\alpha,\beta\right)$  and the severity model \\$X_i(t) \sim S(0.5,1,\gamma,\delta;0)$, then the exact density of the annual loss process can be expressed analytically as a mixture density comprised of $\alpha$-stable components with Poisson mixing weights for $N_t > 0$,
{\small{
\begin{equation}
\begin{split}
f_Z(z) &= \sum_{n=1}^{\infty} \left[ \int_0^{\infty} \exp(-\lambda)\frac{\lambda^n}{n!} \frac{\beta^{\alpha}}{\Gamma(\alpha)}\exp\left(-\beta \lambda \right) \lambda^{\alpha-1}d\lambda \right]  
\left[ \sqrt{\frac{\widetilde{\gamma}_n}{2\pi}}\frac{1}{\left(z-\widetilde{\delta}_n\right)^{3/2}}\exp\left(-\frac{\widetilde{\gamma}_n}{2\left(z-\widetilde{\delta}_n\right)}\right) \right] \times \mathbb{I}\left[\widetilde{\delta}_n<z<\infty\right]\\
&= \sum_{n=1}^{\infty} \frac{(\alpha + n - 1)!}{(\alpha-1)!n!} \left(\frac{\beta}{1+\beta}\right)^{\alpha}\left(\frac{1}{1+\beta}\right)^{n}\left[ \sqrt{\frac{\widetilde{\gamma}_n}{2\pi}}\frac{1}{\left(z-\widetilde{\delta}_n\right)^{3/2}}\exp\left(-\frac{\widetilde{\gamma}_n}{2\left(z-\widetilde{\delta}_n\right)}\right) \right] \times \mathbb{I}\left[\widetilde{\delta}_n<z<\infty\right]\\
\label{GammaPoissMix}
\end{split}
\end{equation}}}
with
\begin{equation*}
\begin{split}
\widetilde{\gamma_n}^{0.5} = \sum_{i=1}^n \gamma_i^{0.5} = n|\gamma|^{0.5}, \; \; \; \; \widetilde{\beta_n} = 1 \; \; \; \;
\widetilde{\delta_n} = \sum_{i=1}^n \delta_i + \tan \frac{\pi}{4}\left(\widetilde{\gamma}_n - \sum_{j=1}^n \gamma_j\right) = n\delta + \tan \frac{\pi}{4}\left(n^2|\gamma| - n\gamma\right),  
\end{split}
\end{equation*}
and $F_Z(0) = \text{Pr}(N_t=0)=\frac{(\beta+1)^{\alpha}}{\Gamma(\alpha)}$ for $N=0$. The exact form of the annual loss cumulative distribution function is also expressible in closed-form,
\begin{equation}
\begin{split}
\Pr\left(Z < z\right) = F_Z(z) &= \sum_{n=1}^{\infty} \frac{(\alpha + n - 1)!}{(\alpha-1)!n!} \left(\frac{\beta}{1+\beta}\right)^{\alpha}\left(\frac{1}{1+\beta}\right)^{n} \text{erfc}\left( \sqrt{\frac{\widetilde{\gamma}_n}{2\left(z-\widetilde{\delta}_n\right)}}\right)\times \mathbb{I}\left[\widetilde{\delta}_n<z<\infty\right] \\
&+ \frac{(\beta+1)^{\alpha}}{\Gamma(\alpha)}. 
\label{GammaPoiss2Mix}
\end{split}
\end{equation}}
Next we present the proof of this result.\\
\textbf{Proof} The result for the convolution between the stable random variables, conditional on $n$, follows from application of Lemma 1, Lemma 2 and Lemma 3. The mixing weight of the compound process for the Poisson-Gamma distribution is derived by considering Bayes Theorem and the conjugacy property of the Poisson-Gamma model from which we know $p(\lambda|n) = \Gamma(\lambda; \alpha + n, \beta + 1)$, hence
\begin{equation}
\begin{split}
\mathbb{P}\text{r}(n) &= \frac{p(n|\lambda)p(\lambda)}{p(\lambda|n)} =\frac{Po(n;\lambda)\Gamma(\lambda;\alpha,\beta)}{\Gamma(\lambda; \alpha + n, \beta + 1)} =\frac{\frac{\lambda^n e^{-\lambda}}{n!}\frac{\beta^{\alpha}}{\Gamma(\alpha)}\lambda^{\alpha-1}e^{-\beta\lambda}}{\frac{(\beta+1)^{\alpha + n}}{\Gamma(\alpha + n)}\lambda^{\alpha + n -1}e^{-(\beta + 1)\lambda}}\\
&= \frac{\Gamma(\alpha+n)}{\Gamma(\alpha)n!}\frac{\beta^{\alpha}}{(1+\beta)^{\alpha + n}}
= \frac{(\alpha + n - 1)!}{(\alpha-1)!n!} \left(\frac{\beta}{1+\beta}\right)^{\alpha}\left(\frac{1}{1+\beta}\right)^{n}. 
\end{split}
\end{equation}
Note, this mixing weight is then a negative binomial probability with $r = \alpha$ and $p = \frac{1}{1+\beta}$ as presented in p.53 of \cite{Gelman}. \qed

Next we will also study the truncation error in these settings for single loss processes. This will be followed by a consideration of some simulation studies comparing the properties of the capital estimation under the binomial, negative binomial and Poisson models and the doubly stochastic models developed. 

\section{Approximation of Infinite Sum Representations by Truncation}
\label{Approx}
In this section we consider analytic expressions for the approximation error of the infinite sum LDA model representations in Theorems 3,4,5,6. To achieve this we will build on the results developed in \cite{Peters10}, which utilsed the properties in Lemma 4. 

\textbf{Lemma 4} \textit{Given a random variable for the severity of a loss $X \sim S(\alpha,\beta,\gamma,\delta;0)$ then as $x \rightarrow \infty$ one can write the limiting tail distribution
\begin{equation}
\begin{split}
P(X>x) &\sim \gamma^{\alpha}c_{\alpha}(1+\beta)x^{-\alpha}, \; \text{as x $\rightarrow \infty$} \\
f_{X}(x|\alpha,\beta,\gamma,\delta;0) &\sim \alpha\gamma^{\alpha}c_{\alpha}(1+\beta)x^{-(\alpha+1)}, \; \text{as x $\rightarrow \infty$.}
\end{split}
\end{equation}
where $c_{\alpha} = \sin\left(\frac{\pi \alpha}{2}\right)\frac{\Gamma(\alpha)}{\pi}$. This result follows from (\cite{Nolan},Theorem 1.12)}

We may now use Lemma 4 to help determine a truncation index for the infinite sums in Theorems 3,4,5 and 6 as show in Theorem 7. To achieve this we will first make an approximation in which we consider the index of the summation to be a continous variable, as demonstrated in for example [Section 3.1] of \cite{Peters09Mod}. In addition, in this paper we will work with the analytic compound process expression for the tail asymptotic, and ensure that after truncation, the error in the finite expansion of the tail asympotic is bounded to a desired precision. More precisely, the asymptotic tail probability of the compound process will be established to have a desired approximation error under the truncation derived.
This is not unique, and equally we could have chosen other distributional properties such as the median that was used in \cite{Peters10}.

\textbf{Theorem 7} \textit{Given the LDA structure in which the frequency is distributed generically according to $N(t) \sim f(\cdot)$ and the severity model $X_i(t) \sim S(0.5,1,\gamma,\delta;0)$, the exact expressions for the tail asymptotic of the resulting infinite mixture is given by,
\begin{equation*}
\begin{split}
\text{Pr}\left(Z(t)>z_q\right) &\sim 2x^{-0.5}c_{0.5} \sum_{n=1}^{\infty}Pr(N(t)=n) \widetilde{\gamma}_n^{0.5} = C_{\text{Tail}}\sum_{n=1}^{\infty} W_n 
\end{split}
\end{equation*}
where $z_q$ is an upper tail quantile of the annual loss distribution and it will be assumed for simplicity that $\widetilde{\gamma_n}^{0.5} = \sum_{i=1}^n \gamma_i^{0.5} = n|\gamma|^{0.5}$. These results allows us to determine a unique maximum of $\frac{d}{dn}\log(W_n) = 0$ corresponding to the term $n$ with the maximum contribution to the LDA compound processes tail probability for each model. Note this can be determined trivially for each model by searching for a sign change in the function for each integer value of $n$ and so is very efficient to solve since only requires integer search. The specified equations for each model are given as follows:\newline
\underline{Poisson-$\alpha$-Stable.}\newline
Under this model we have that $W_n = |\gamma|^{0.5}\exp(-\lambda)\frac{\lambda^n}{(n-1)!}$ which results in the function $\frac{d}{dn}\log(W_n) = 0$ producing $\log(\lambda) - \log(n) - \frac{1}{2n}=0$.\newline
\underline{Doubly Stochastic Poisson-Gamma-$\alpha$-Stable.}\newline
Under this model we have that $W_n = |\gamma|^{0.5}
\frac{(\alpha + n - 1)!}{(\alpha-1)!n!} \left(\frac{\beta}{1+\beta}\right)^{\alpha}\left(\frac{1}{1+\beta}\right)^{n}$ which results in the function $\frac{d}{dn}\log(W_n) = 0$ producing $\log(n + \alpha - 1) + \frac{n}{(n + \alpha - 1)} + \frac{(\alpha - 0.5)}{(n + \alpha - 1)} -\log(n) - 1 - \frac{1}{2n}  - \log(\beta + 1) =0$. \newline
\underline{negative binomial-$\alpha$-Stable.}\newline
Under this model we have that $W_n = |\gamma|^{0.5} \frac{(n+r-1)!}{(n+r-1-n)!(n)!}\left(1-p\right)^r \left(p\right)^{n}$ which results in the function $\frac{d}{dn}\log(W_n) = 0$ producing $\log(n+r-1) + \frac{n}{n+r-0.5} +   \frac{r-0.5}{n+r-1} - 1 - \log(n) - \frac{1}{2n} + \log(p) = 0$. \newline
\underline{Doubly Stochastic negative binomial- Beta-$\alpha$-Stable.}\newline
Under this model we have that $W_n = |\gamma|^{0.5}\frac{(n+r-1)!p^{(\alpha + r -1)}B(\alpha + rn, \beta + n)}{(r-1)!n!p^{(\alpha + rn -1)}B(\alpha,\beta)}$which results in the function $\frac{d}{dn}\log(W_n) = 0$ producing,
\begin{equation}
\begin{split} 
&\log(n+r) + \frac{(n+r+0.5)}{(n+r)} + r\log(\alpha + rn) + \frac{r(\alpha + rn-0.5)}{(\alpha + rn)} + \log(\beta + n) + \frac{(\beta + n-0.5)}{(\beta + n)}\\ 
&- (r+1)\log(\alpha + rn+\beta + n) - \frac{(r+1)(\alpha + rn + \beta + n -0.5)}{(\alpha + rn+\beta + n)} - \log(n+1) - \frac{(n+1+0.5)}{(n+1)} - r\log(p) = 0.
\end{split}
\end{equation}
We can now find $W_L < W_0 < W_U$ such that the truncated sum approximation is sufficiently accurate for use in the expression of the annual loss distribution of the compound process. Hence, we can bound the number of terms in the mixture representations to not change by a precision amount $\epsilon$. This is expressed analytically as 
\begin{equation}
f_Z(z) = \sum_{n=N_L}^{N_U} Pr(N_t = n) \left[ \sqrt{\frac{\widetilde{\gamma}_n}{2\pi}}\frac{1}{\left(x-\widetilde{\delta}_n\right)^{3/2}}\exp\left(-\frac{\widetilde{\gamma}_n}{2\left(x-\widetilde{\delta}_n\right)}\right) \right]
\label{PoissMix}
\end{equation}
with $N_L$ and $N_U$ are determined from $\max(W_L \leq e^{-37}W_{0},1)$ and $W_U \leq e^{-37}W_{0}$.}
The derivation of these results are found in Appendix \ref{Append2}. We will demonstrate simulated examples for these results on the truncation in the following results and discussion section.

\section{Simulation Studies}
In this section we undertake several simulation studies, for all studies we will utilise the parameter settings in Table \ref{SimulationSettings}. In addition, the calculated values for $N_L$ and $N_U$ are provided for each of these models, utlising the results derived in Theorem 7 in Section \ref{Approx}. For each of the frequency distributions, we considered two settings, low frequency and high frequency. In addition, all simulations were completed in Matlab with vectorised code on an SGI cluster in which the compute node utilised was an Altix XE320 with two Intel Xeon X5472 (quad core 3.0GHz) CPU's, the source code available at: \vspace{-0.5cm}
\begin{verbatim}http://web.maths.unsw.edu.au/~peterga/index_files/QRSLab/TechnicalReports.html
\end{verbatim}

\begin{table}[!h]
\begin{tabular}{|c|c|c|c|c|c|c|c|c|}\hline
LDA Model & \multicolumn{4}{|c|}{Frequency Parameters} & \multicolumn{2}{|c|}{Severity Parameters} & \multicolumn{2}{|c|}{Truncation}\\ 
 														& \multicolumn{1}{|c}{$\lambda$} &\multicolumn{1}{c}{n} 	& \multicolumn{1}{c}{p} 			& \multicolumn{1}{c|}{r} & \multicolumn{1}{c}{$\gamma$}& \multicolumn{1}{c|}{$\delta$} & \multicolumn{1}{c}{$N_L$}& \multicolumn{1}{c|}{$N_U$}\\ \hline \hline
\multicolumn{9}{|c|}{Low Frequency Example} \\\hline
binomial-Levy 							&		-				& 12 	& 0.1 		& -	 	& 0.01 & 0 & 0& M\\
binomial-Beta-Levy 					&		-			  & 12 	& Be(1,5) & -		& 0.01 & 0 & 0& M\\
negative binomial-Levy 			&  	-				& - 	& 0.1			& 2 	& 0.01 & 0 & 0& 19\\
negative binomial-Beta-Levy & 	-				&	-		& Be(1,5) & 2 	& 0.01 & 0 & 0& 20\\
Poisson-Levy 								&	0.1				& -		& -				& - 	& 0.01 & 0 & 0& 11\\	
Poisson-Gamma-Levy					& Ga(1,0.1)	& -		& -				& - 	& 0.01 & 0 & 0& 171\\ \hline				
\multicolumn{9}{|c|}{High Frequency Example} \\\hline
binomial-Levy 							&		-				& 12 	& 0.6 		& -		& 0.01 & 0 & 0& M\\
binomial-Beta-Levy 					&		-			  & 12 	& Be(5,1) & -		& 0.01 & 0 & 0& M\\
negative binomial-Levy 			&  	-				& - 	& 0.6			& 10 	& 0.01 & 0 & 0& 120\\
negative binomial-Beta-Levy & 	-				&	-		& Be(5,1) & 10 	& 0.01 & 0 & 0& 169\\
Poisson-Levy 								&	10				& -		& -				& - 	& 0.01 & 0 & 0& 49\\	
Poisson-Gamma-Levy					& Ga(1,10)	& -		& -				& - 	& 0.01 & 0 & 0& 171\\	\hline				
\end{tabular}
\label{SimulationSettings}
\caption{Parameter settings for the models utilised in the simulations.}
\end{table}

\subsection{Analysis of Model Estimation under Exact Truncated Solutions versus Monte Carlo}
In this section we first illustrate that the derived annual loss compound process distributions in Theorems 1 to Theorem 6 are correct. This is achieved by undertaking a simulation study in which we simulate from each of the models developed first via a standard Monte Carlo procedure, utilising $T=1,000,000$ annual simulated years for each LDA model. We record the simulation time as well as the estimated Monte Carlo error obtained by blocking the simulated years into sets of 50,000 samples, used to calculated the standard error in estimates obtained from the simulated annual losses. We then estimate the empirical distribution function and plot this along with the estimatd standard error from the Monte Carlo simulation. Next we evaluate each of the models using truncations of the exact expressions derived in Theorem 1 through to Theorem 6, where we utilise $N_L = 1$ and $N_U = 1,000$. The evaluations are done at 200 points from $Z=1$ to $Z=200$ which for the parameter settings in Table \ref{SimulationSettings} covered sufficient range for the evaluation to include quantiles representing regulatory capital VaR's at $99.5\%$. We then plot the evaluated distribution functions versus the Monte Carlo estimated empirical distributions functions for each of the compound process models. 

\begin{figure}[!ht]
\includegraphics[width = 0.5\textwidth, height = 5cm]{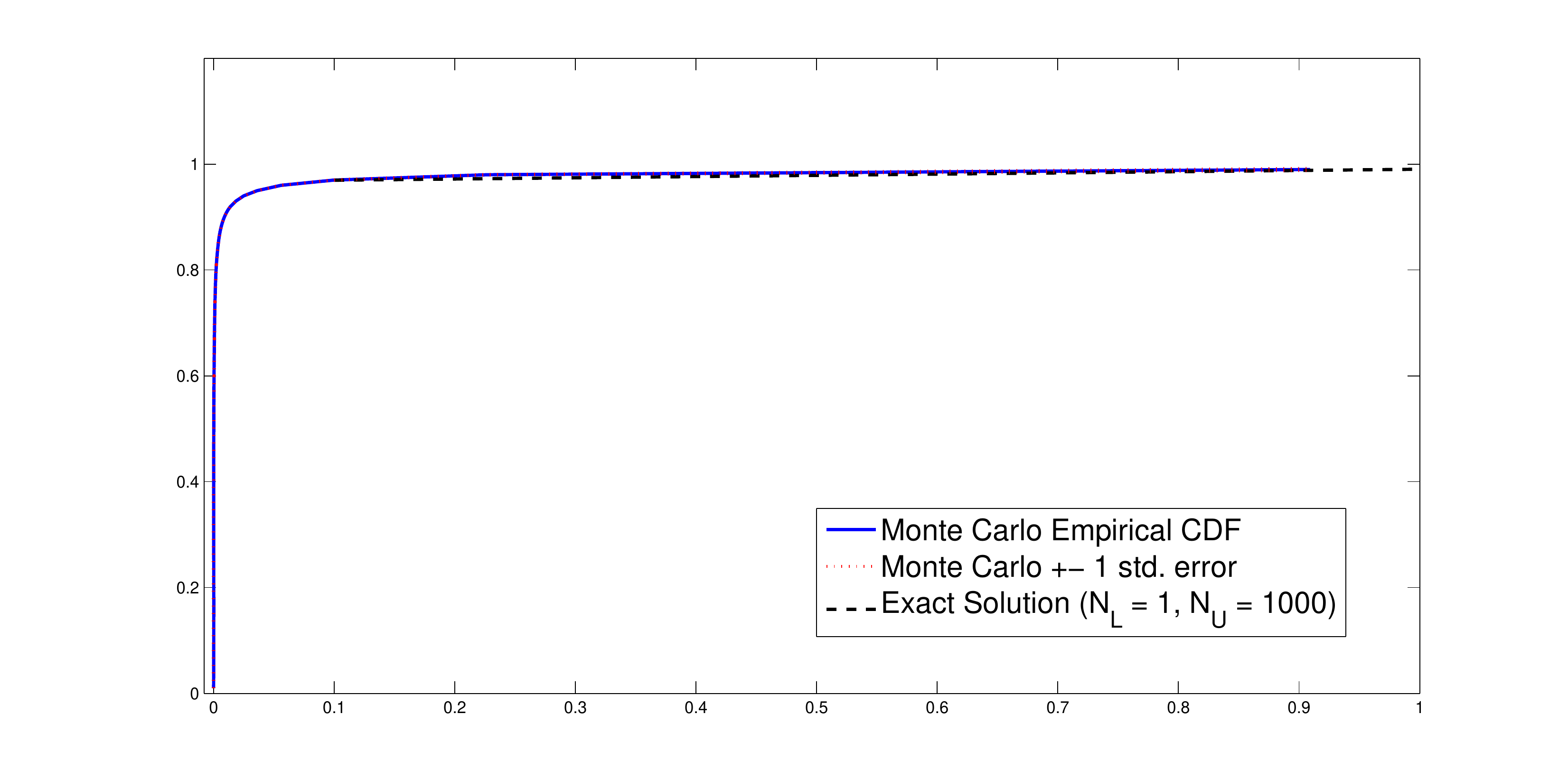}
\includegraphics[width = 0.5\textwidth, height = 5cm]{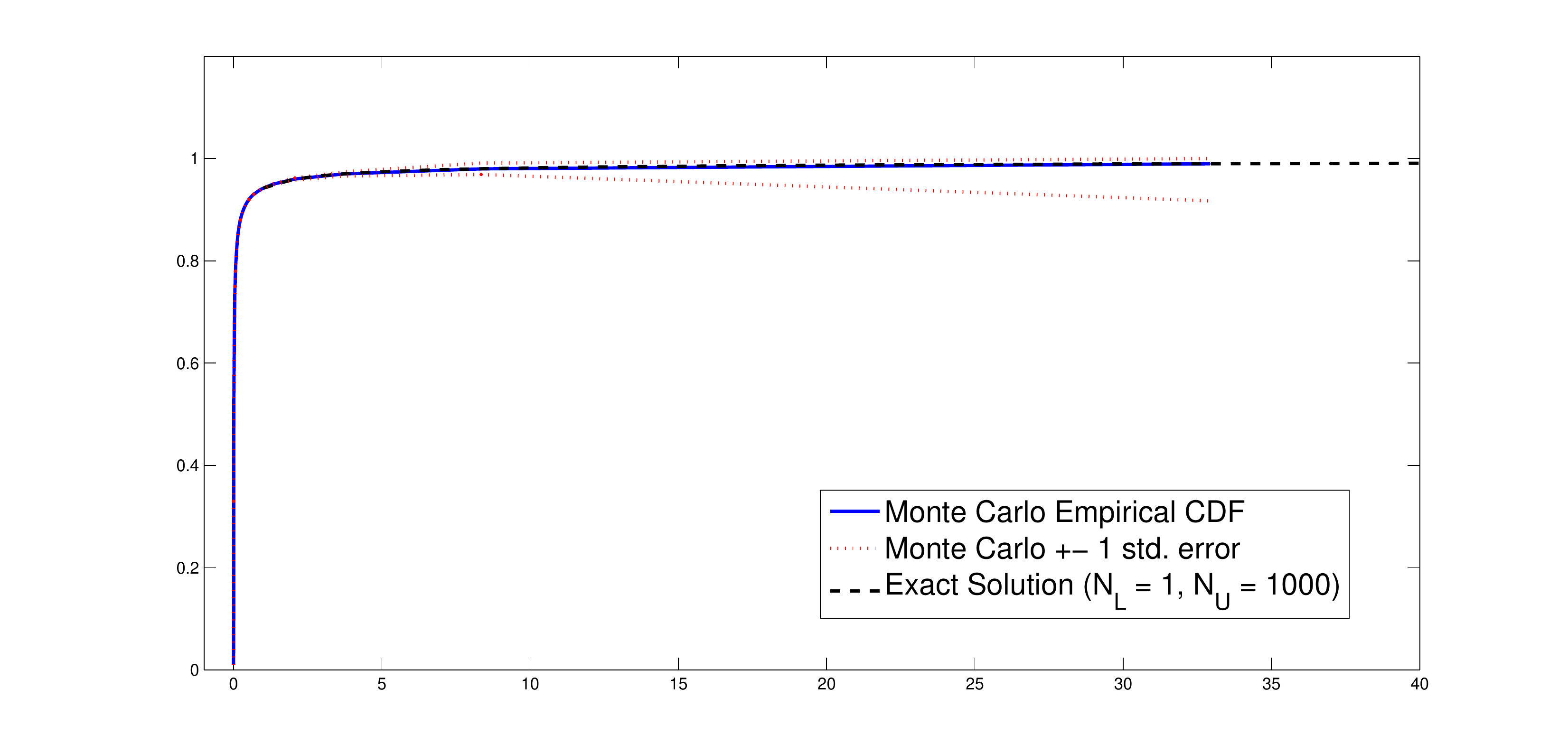}
\includegraphics[width = 0.5\textwidth, height = 5cm]{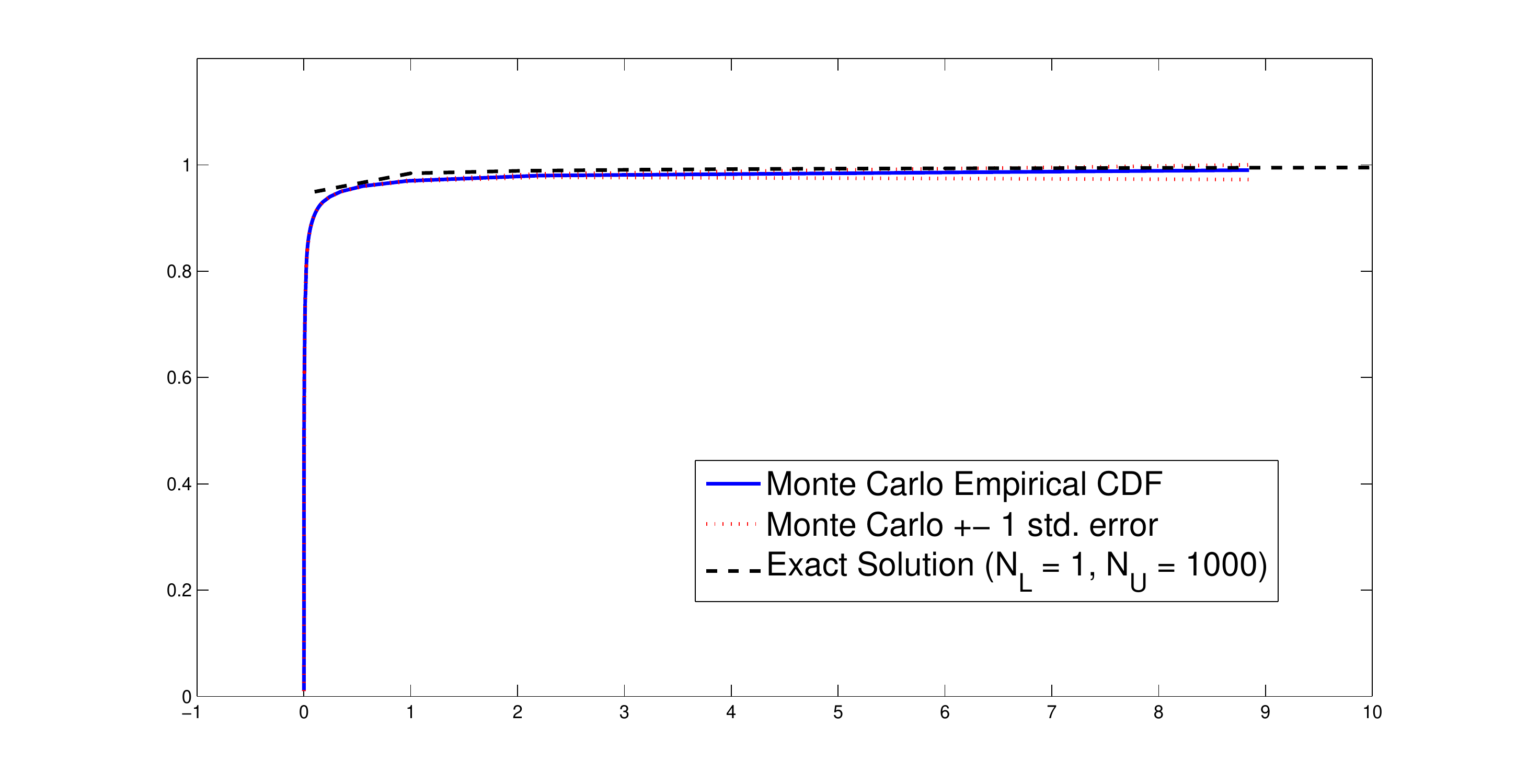}
\includegraphics[width = 0.5\textwidth, height = 5cm]{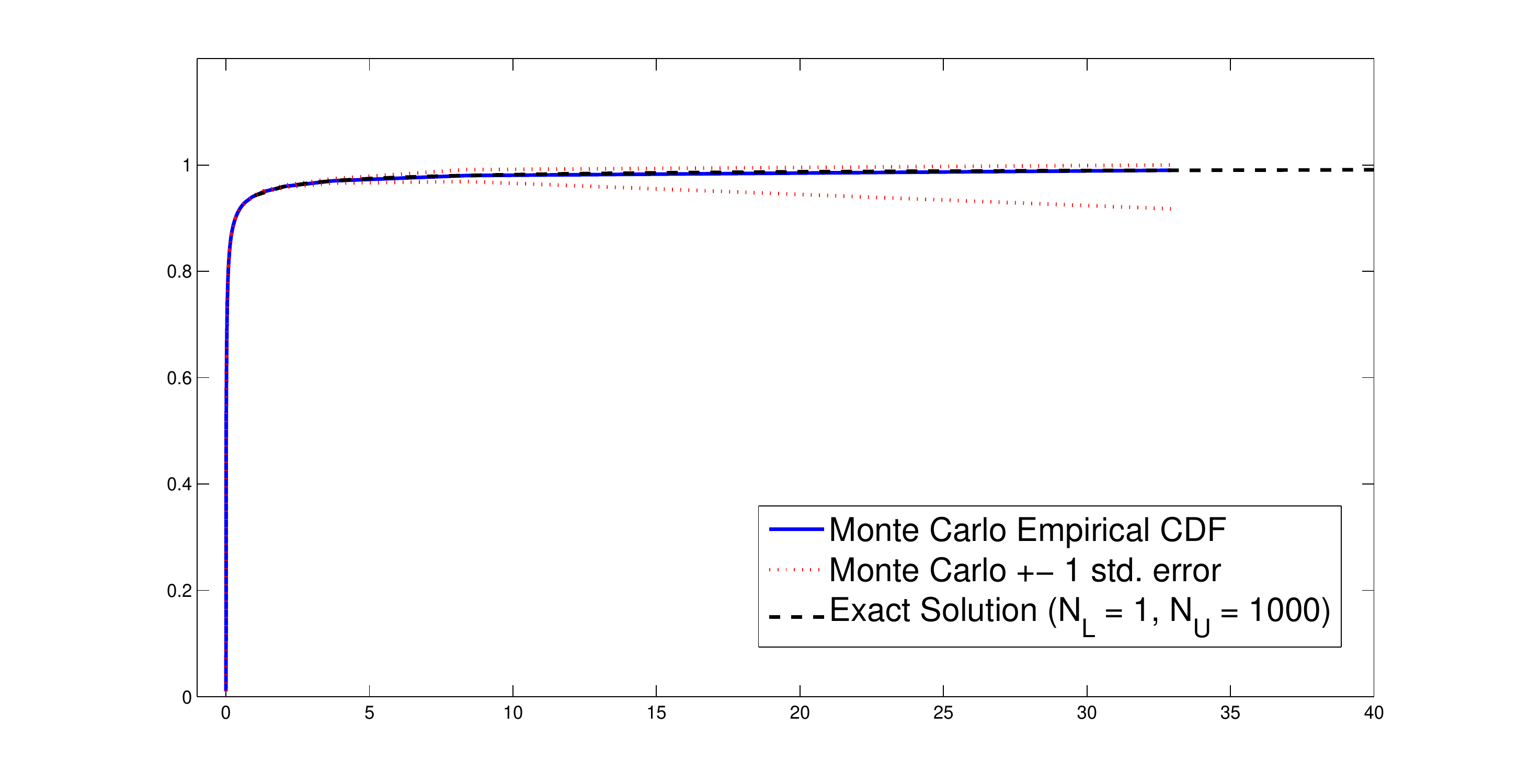}
\caption{Comparison of epirical CDF from $T=1,000,000$ simulated annual years (dashed lines), versus Truncated $N_U = 1000$ Exact Expressions (solid line). \textbf{Top Left} - Low frequency Poisson-Levy model  \textbf{Top Right} - High frequency Poisson-Levy model \textbf{Bottom Left} - Low frequency Poisson-Gamma-Levy model  \textbf{Bottom Right} - High frequency Poisson-Gamma-Levy model}
\label{fig:binomialLevy}
\end{figure}

In Figure \ref{fig:binomialLevy} we present the results of this simtulation for the Annual loss compound process with Levy severity distribution and the binomial and the doubly stochastic binomial-Beta models derived in Theorem 1 and Theorem 2. The distribution functions are presented for both the low and high frequency examples.

\begin{figure}[!ht]
\includegraphics[width = 0.5\textwidth, height = 5cm]{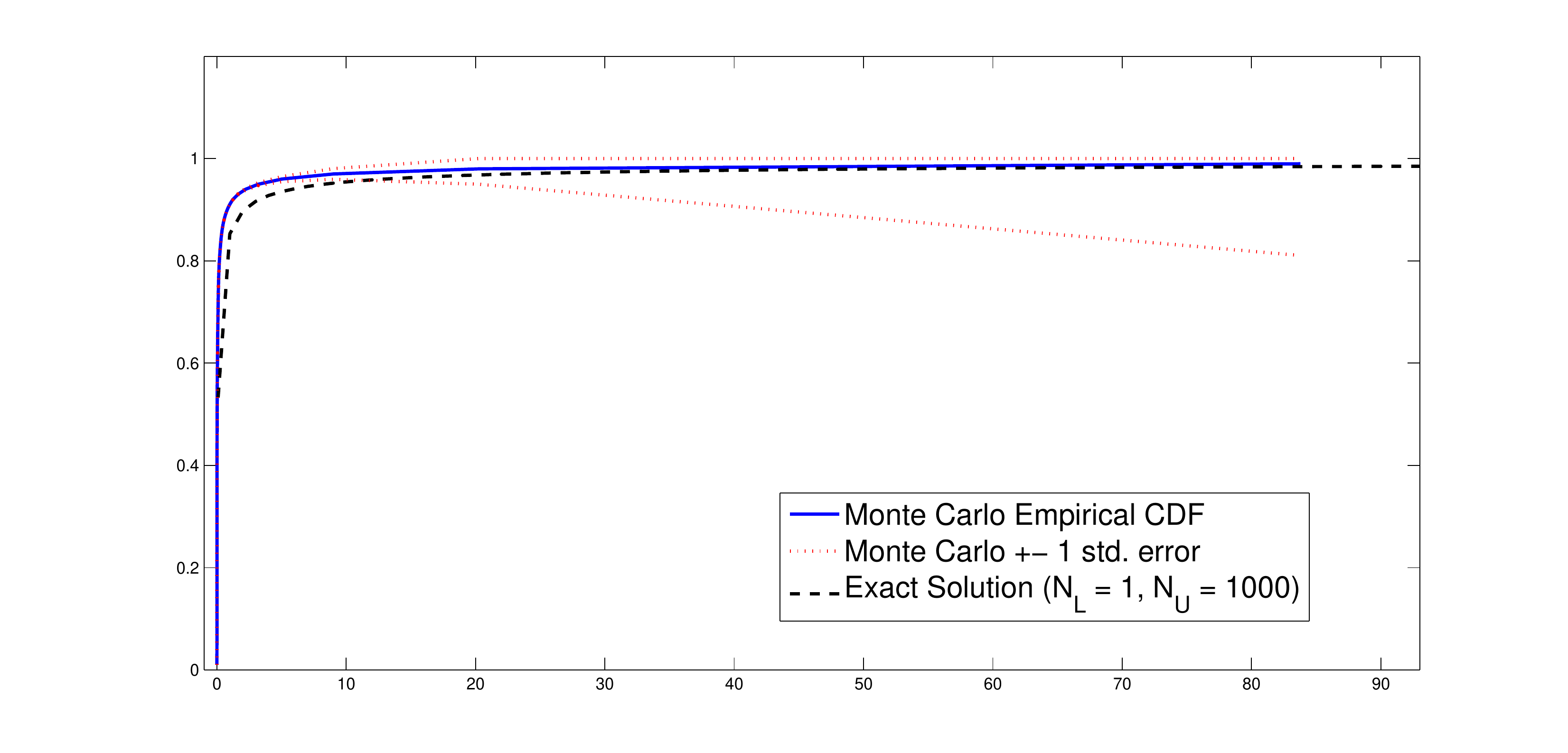}
\includegraphics[width = 0.5\textwidth, height = 5cm]{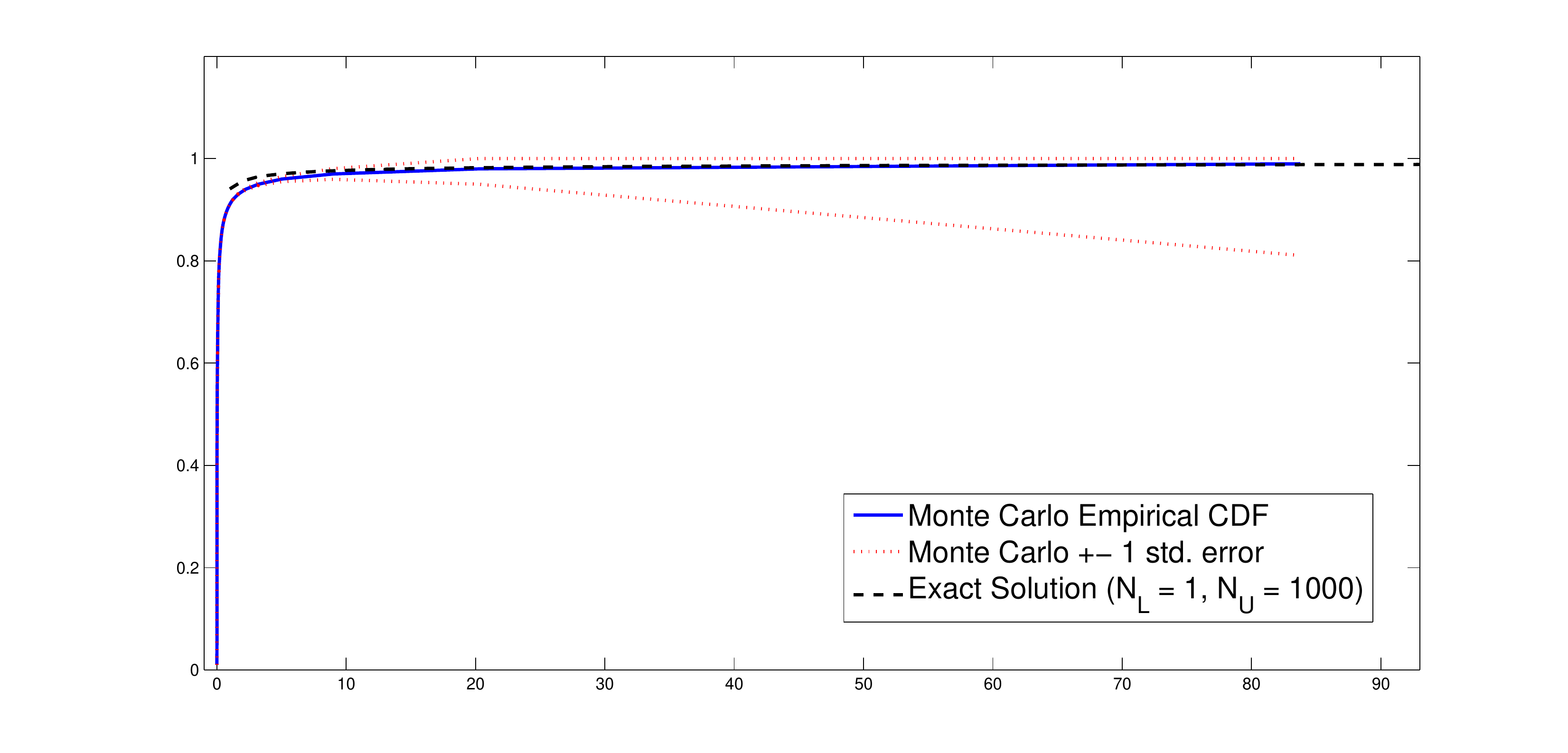}
\includegraphics[width = 0.5\textwidth, height = 5cm]{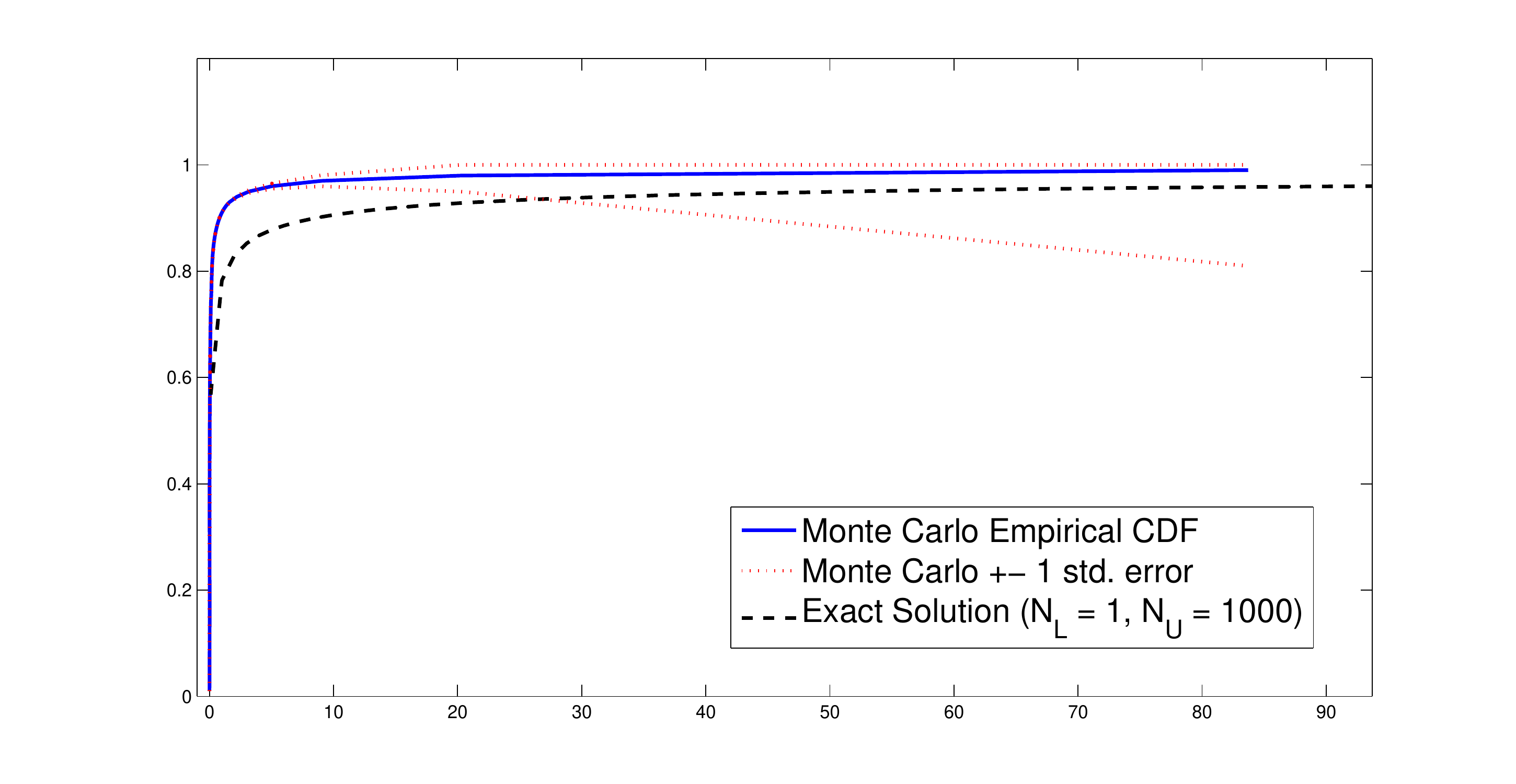}
\includegraphics[width = 0.5\textwidth, height = 5cm]{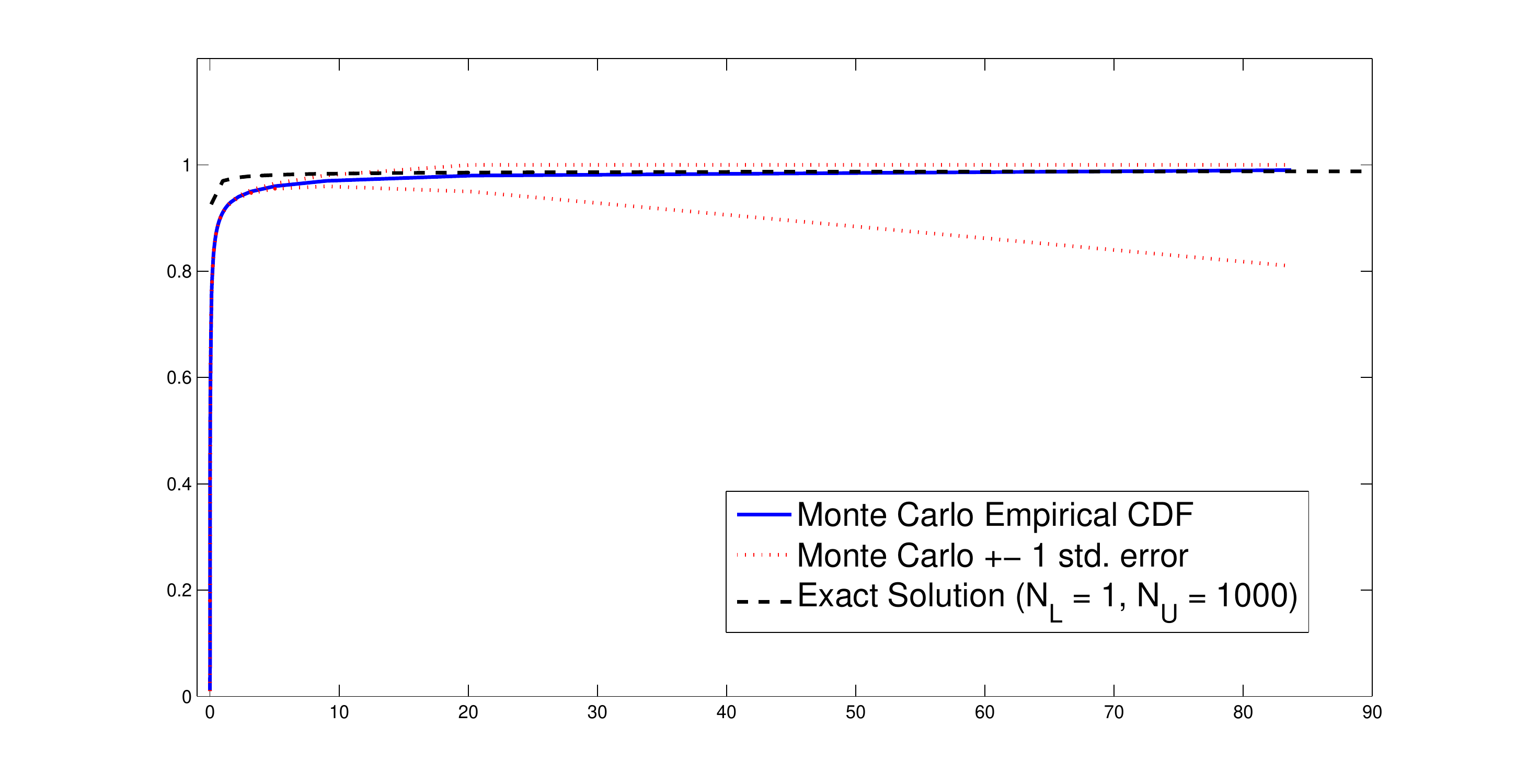}
\caption{Comparison of epirical CDF from $T=1,000,000$ simulated annual years (dashed lines), versus Truncated $N_U = 1000$ Exact Expressions (solid line). \textbf{Top Left} - Low frequency Poisson-Levy model  \textbf{Top Right} - High frequency Poisson-Levy model \textbf{Bottom Left} - Low frequency Poisson-Gamma-Levy model  \textbf{Bottom Right} - High frequency Poisson-Gamma-Levy model}
\label{fig:negativebinomialLevy}
\end{figure}

In Figure \ref{fig:negativebinomialLevy} we present the results of this simtulation for the Annual loss compound process with Levy severity distribution and the negative binomial and the doubly stochastic negative binomial-Beta models derived in Theorem 5 and Theorem 6. The distribution funcstions are presented for both the low and high frequency examples.

\begin{figure}[!ht]
\includegraphics[width = 0.5\textwidth, height = 5cm]{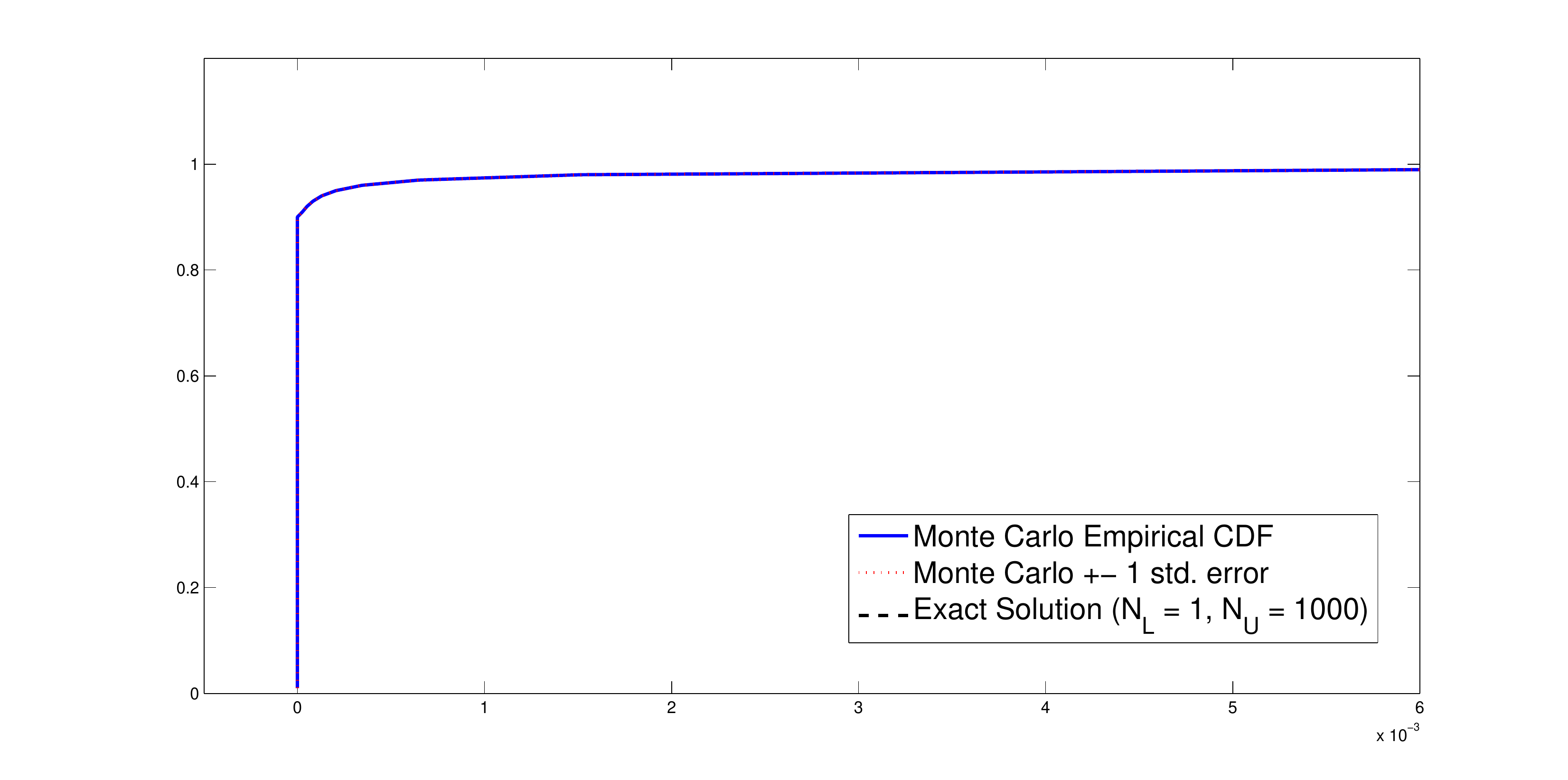}
\includegraphics[width = 0.5\textwidth, height = 5cm]{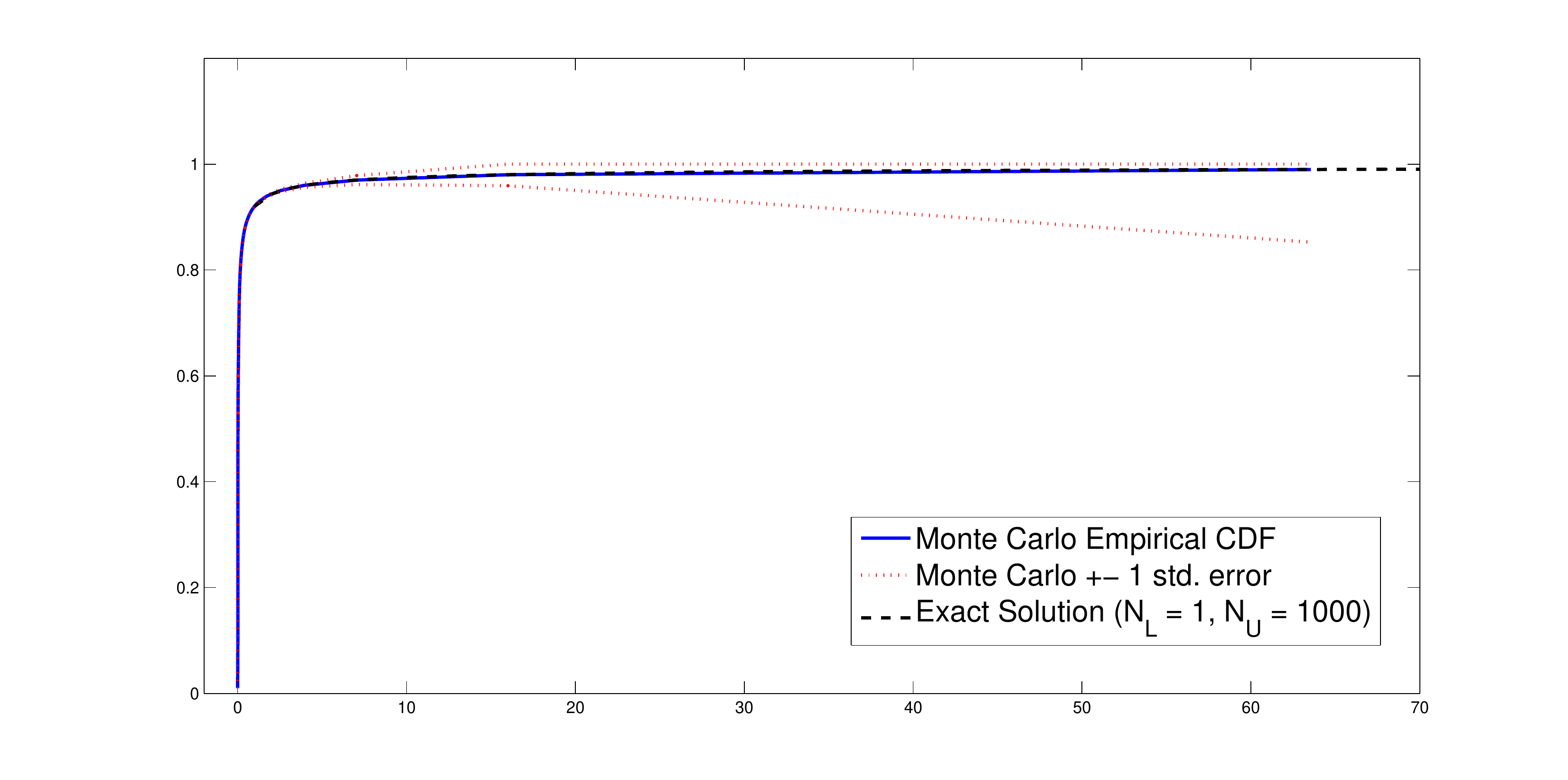}
\includegraphics[width = 0.5\textwidth, height = 5cm]{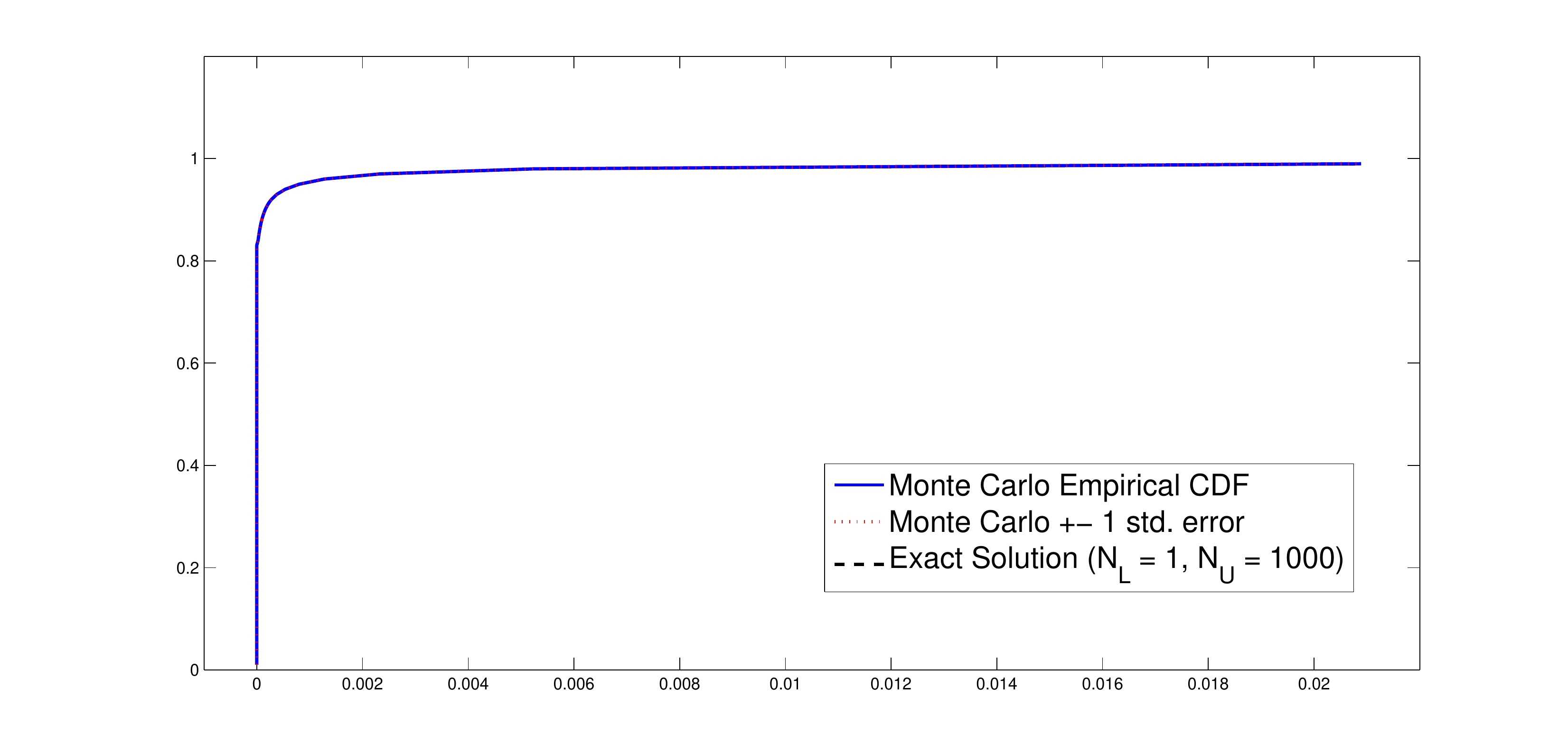}
\includegraphics[width = 0.5\textwidth, height = 5cm]{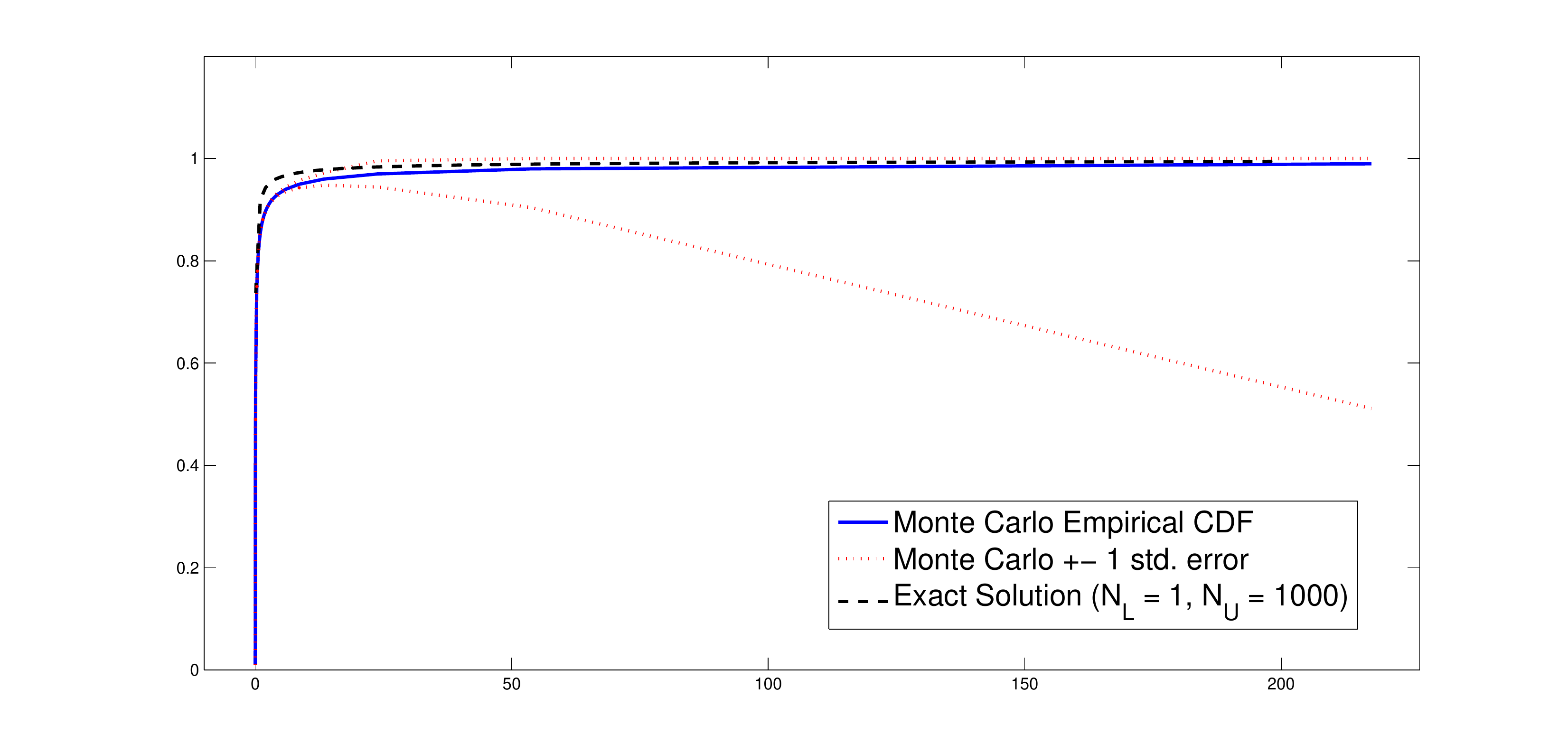}
\caption{Comparison of epirical CDF from $T=1,000,000$ simulated annual years (dashed lines), versus Truncated $N_U = 1000$ Exact Expressions (solid line). \textbf{Top Left} - Low frequency Poisson-Levy model  \textbf{Top Right} - High frequency Poisson-Levy model \textbf{Bottom Left} - Low frequency Poisson-Gamma-Levy model  \textbf{Bottom Right} - High frequency Poisson-Gamma-Levy model}
\label{fig:PoissonLevy}
\end{figure}

In Figure \ref{fig:PoissonLevy} we present the results of this simtulation for the Annual loss compound process with Levy severity distribution and the Poisson and the doubly stochastic Poisson-Gamma models derived in Theorem 5 and Theorem 6. The distribution funcstions are presented for both the low and high frequency examples.

\section{Analysis of Truncation Error and Computational Saving versus Monte Carlo}
Monte Carlo techniques are widely used in the generation of compound processes and especially in the estimation of the annual loss distribution and the associated risk measures of Value at Risk (VaR) and Expected Shortfall (ES). In this section we fist study the truncation error when applied to the models derived in Theorem 1 through to Theorem 6 and then we demonstrate the accuracy of the derived models under the estimated truncations $N_L$ and $N_U$. Finally, we compared the simulation times for each method when we very conservatively set $N_L = 1$ and $N_U = 1,000$, well in excess of the required estimates truncations for these examples as shown below.

\begin{figure}[!ht]
\includegraphics[width = 1\textwidth, height = 5cm]{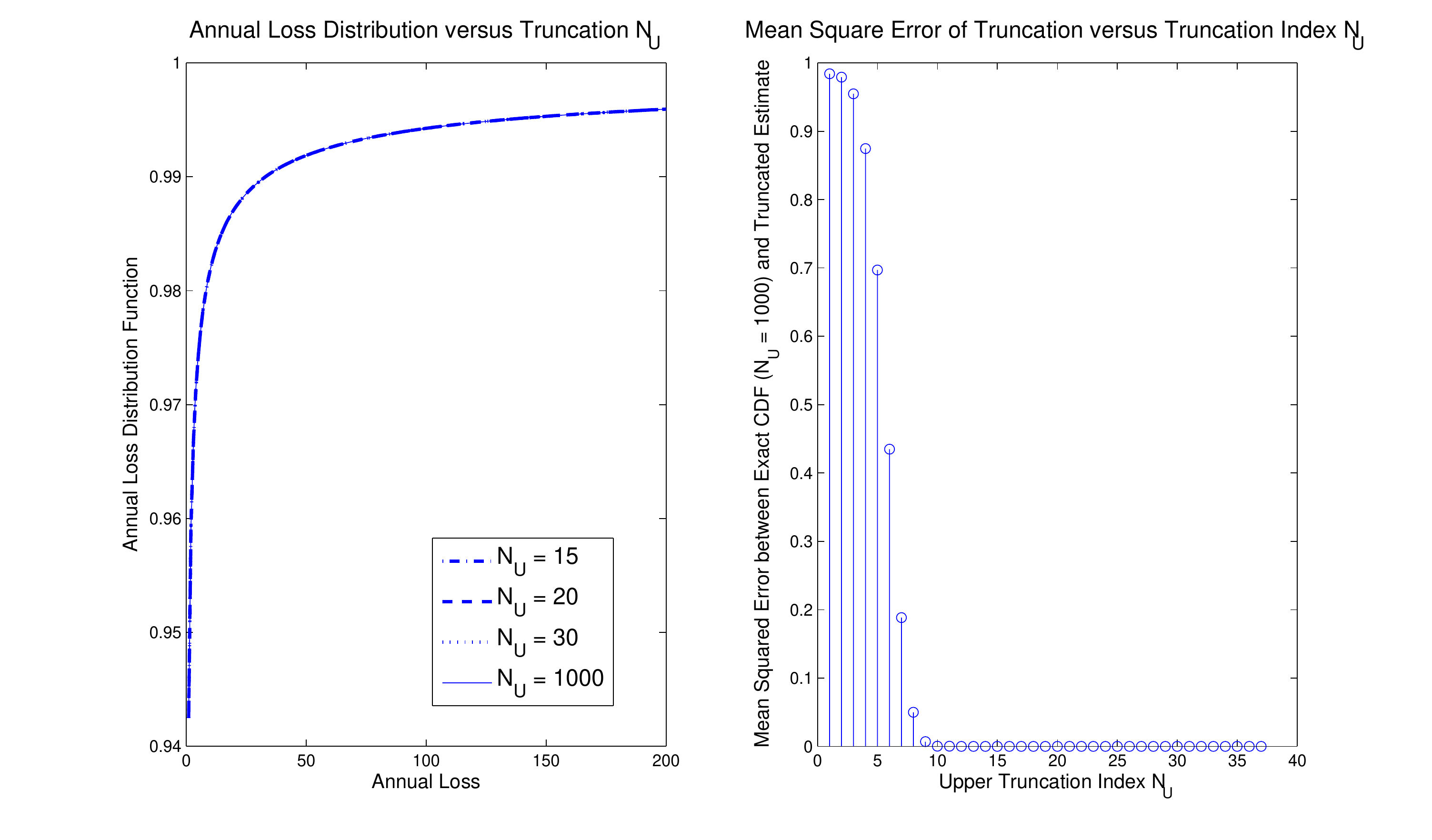}
\includegraphics[width = 1\textwidth, height = 5cm]{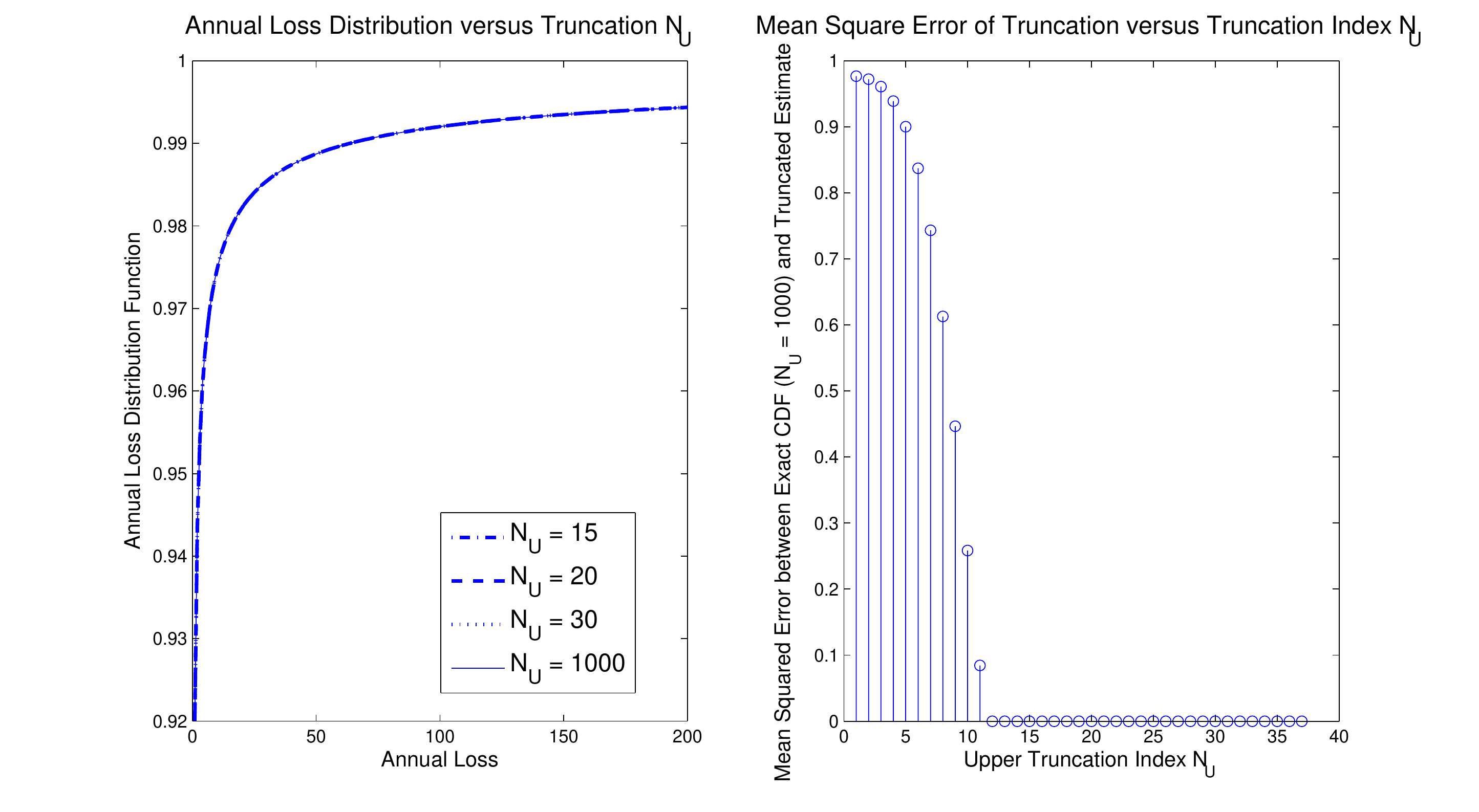}
\caption{Comparison of high frequency Poisson-Levy LDA model epirical CDF versus Truncation Threshold $N_U$ - Exact Expression ($N_U=1000$) (solid line). \textbf{Left} - Estimated CDF versus $N_U$ \textbf{Right} - Mean Square Error versus $N_U$}
\label{fig:BinLevyError}
\end{figure}

\begin{figure}[!ht]
\includegraphics[width = 1\textwidth, height = 5cm]{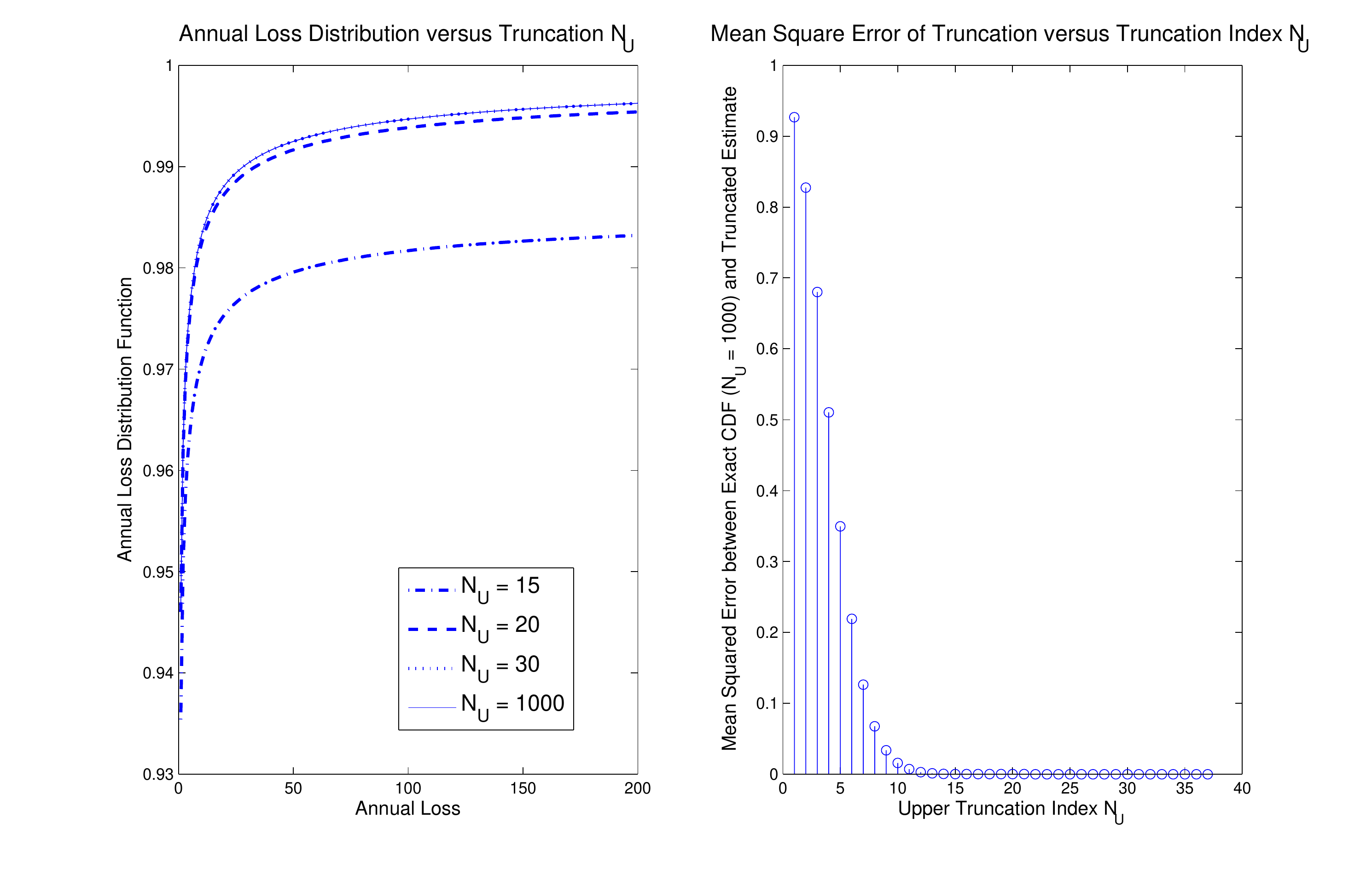}
\includegraphics[width = 1\textwidth, height = 5cm]{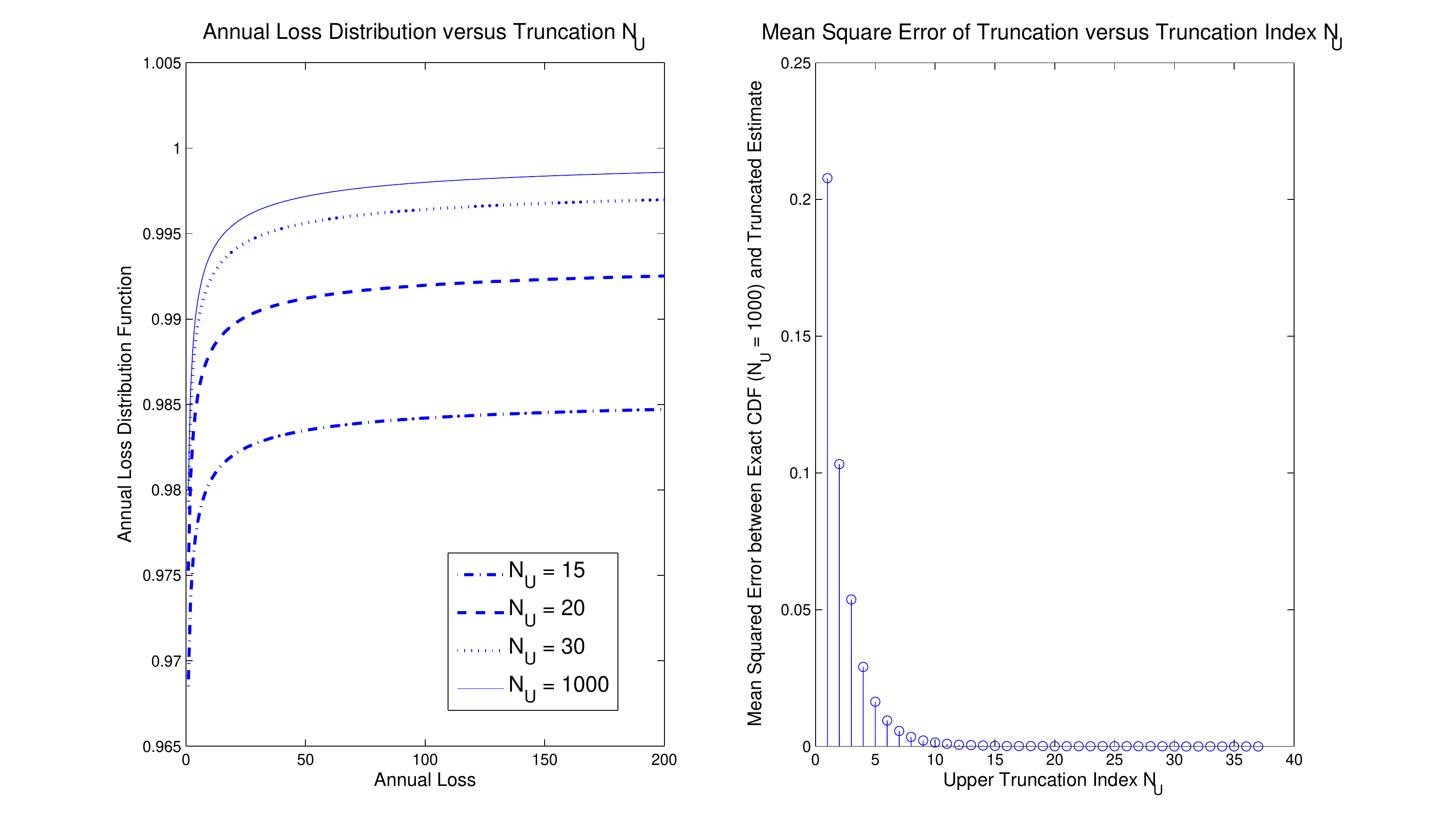}
\caption{Comparison of high frequency Poisson-Levy LDA model epirical CDF versus Truncation Threshold $N_U$ - Exact Expression ($N_U=1000$) (solid line). \textbf{Left} - Estimated CDF versus $N_U$ \textbf{Right} - Mean Square Error versus $N_U$}
\label{fig:NegBinLevyError}
\end{figure}

\begin{figure}[!ht]
\includegraphics[width = 1\textwidth, height = 5cm]{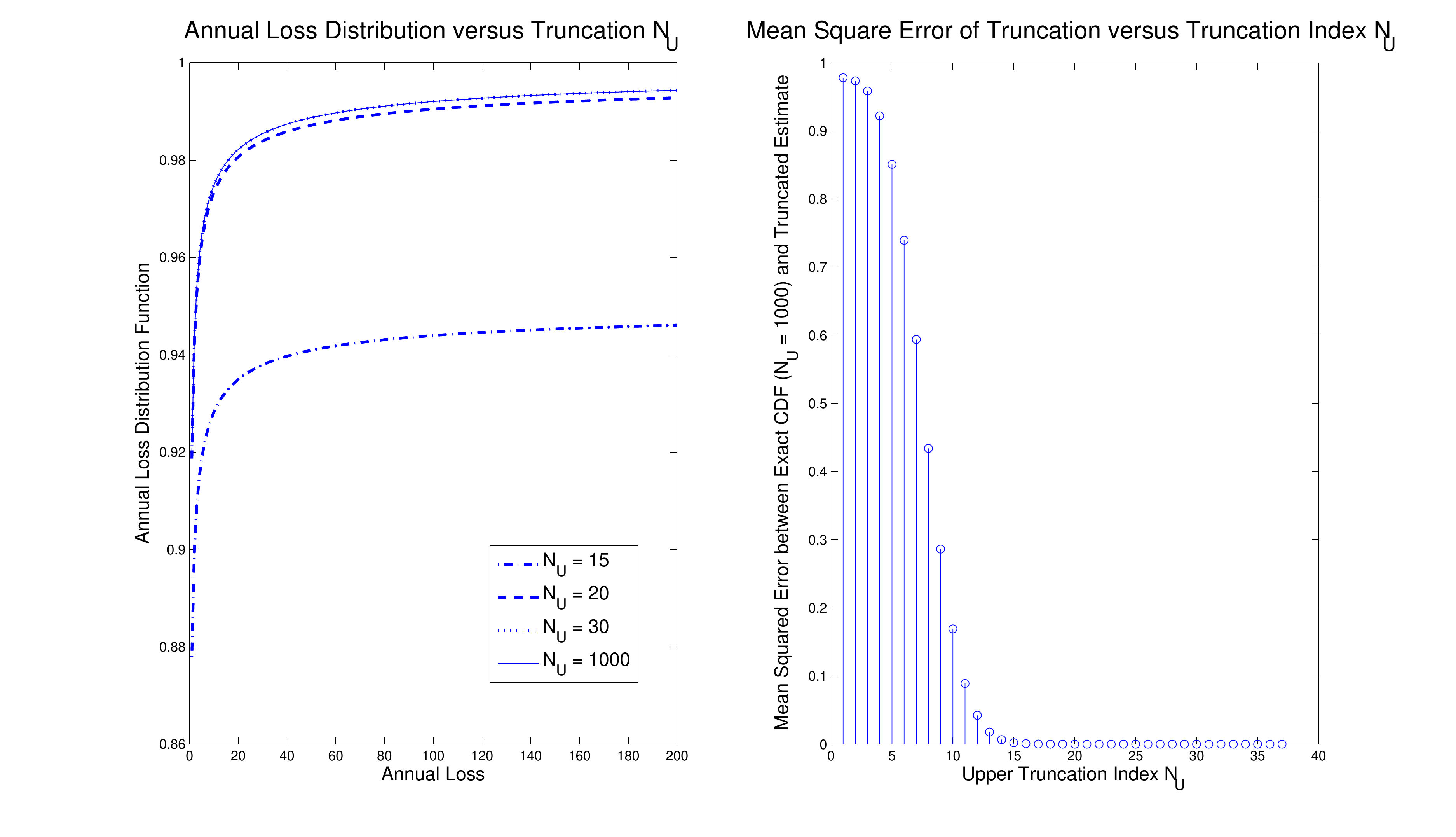}
\includegraphics[width = 1\textwidth, height = 5cm]{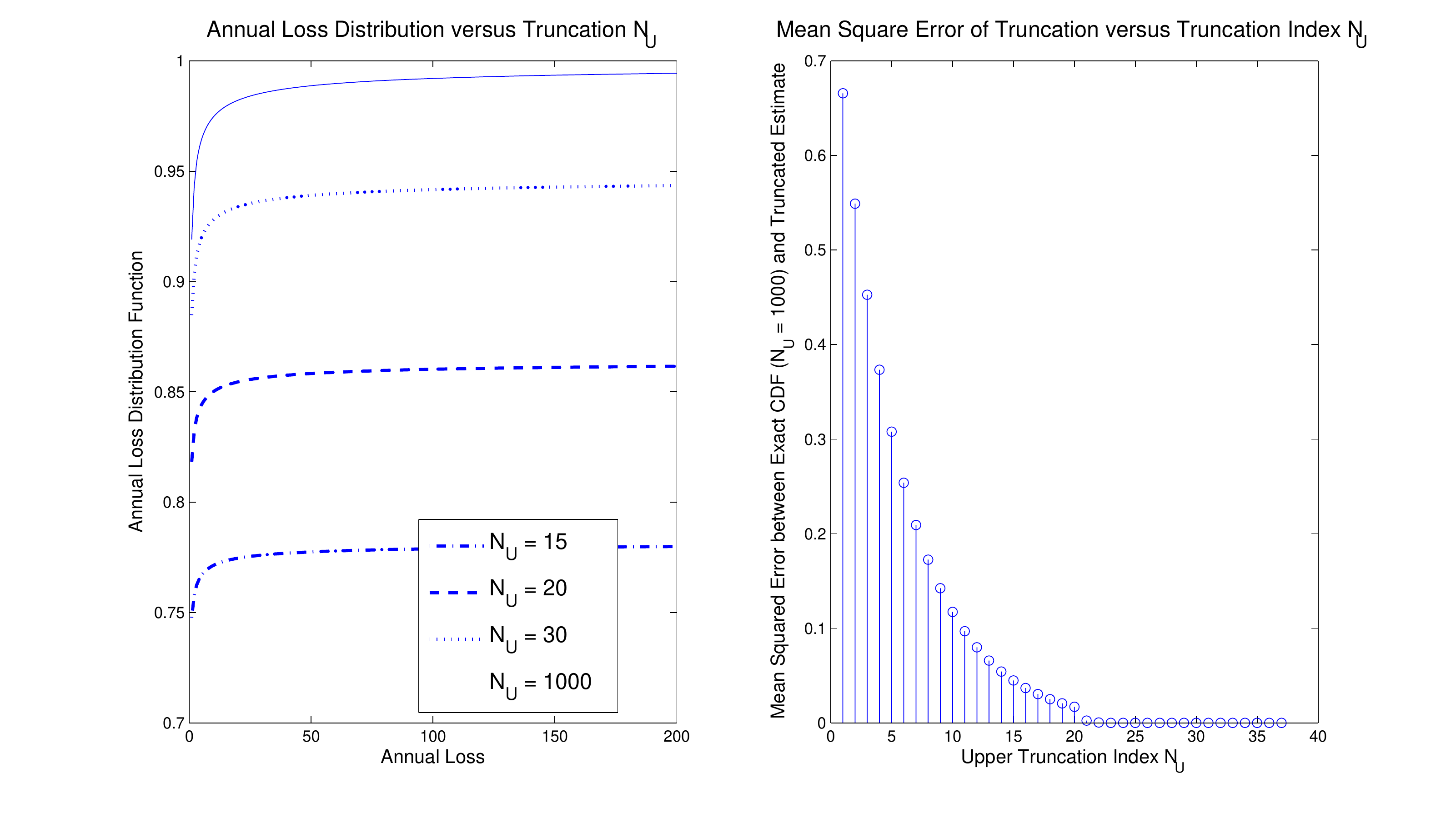}
\caption{Comparison of high frequency Poisson-Levy LDA model epirical CDF versus Truncation Threshold $N_U$ - Exact Expression ($N_U=1000$) (solid line). \textbf{Left} - Estimated CDF versus $N_U$ \textbf{Right} - Mean Square Error versus $N_U$}
\label{fig:PoissonLevyError}
\end{figure}

We demonstrate the estimated CDF and MSE for different levels of truncation $N_U \in \left\{1,2,\ldots,1,000\right\}$ in Figure \ref{fig:BinLevyError} for the binomial and doubly stochastic binomial-Beta examples. Figure \ref{fig:NegBinLevyError} shows the negative binomial and doubly stochastic negative binomial-Beta examples and Figure \ref{fig:PoissonLevyError} shows the Poisson and doubly stochastis Poisson-Gamma examples.

These results demonstrate that in all cases, the mean squared error falls rapidly for small numbers of upper truncation index $N_U$. This confirms the utility of utilising these truncations, which are extremely efficient simulation wise, where in Table \ref{SimulationSettingsTruncation} we see that even when using an excessive number of terms $N_U = 1,000$ we make very significant computational savings and accuracy improvements over the Monte Carlo equivalent simulations.

\begin{table}
\begin{tabular}{|ccc|}\hline
Model & Exact Truncated ($N_L = 1, N_U = 1,000$) & Monte Carlo ($T=1,000,000$)\\ \hline \hline
\multicolumn{3}{|c|}{Low Frequency Examples} \\
binomial-Levy 									& 6.1 (sec) & 10 (min)\\
binomial-Beta-Levy 							& 1.2 (sec) & 8 (min)\\
negative binomial-Levy 					& 4.1 (sec) & 4 (min)\\
negative binomial-Beta-Levy 		& 1.2 (sec) & 3 (min)\\
Poisson-Levy 										& 1.2 (sec) & 8 (min)\\
Poisson-Gamma-Levy 							& 1.2 (sec) & 7 (min) \\
\multicolumn{3}{|c|}{High Frequency Examples} \\
binomial-Levy 									& 6.1 (sec) & 7 (min) \\
binomial-Beta-Levy 							& 1.2 (sec) & 7 (min) \\
negative binomial-Levy 					& 4.2 (sec) & 7 (min) \\
negative binomial-Beta-Levy 		& 1.2 (sec) & 5 (min) \\
Poisson-Levy 										& 1.2 (sec) & 10 (min)\\
Poisson-Gamma-Levy 							& 1.2 (sec) & 6 (min)\\ \hline
\end{tabular}
\caption{\label{SimulationSettingsTruncation} Simulation times for exact solutions versus Monte Carlo estimates.}
\end{table}

Next we present a general result which provides a closed form distributional result for the combined loss process for multiple risk processes, each under one of the LDA model structures presented.

\section{Multiple Risk Process Aggregation}
In this section we demonstrate how to obtain the distribution of the combined loss processes under each of the models developed by presenting a general closed form result for the aggregation of two risk processes which is trivially extended to any number of risk processes. Then we illustrate this result with an example using two Poisson processes LDA models.

\textbf{Theorem 8} \textit{The distribution of the annual loss process represented by multiple risks, eg. $j \in \left\{1,2\right\}$, in a LDA compound process structure in which the frequency is $N^{(j)}(t) \sim F(\theta^{(j)})$ and the severity model $X_i^{(j)}(t) \sim S(0.5,1,\gamma^{(j)},\delta^{(j)};0)$, can be expressed analytically as a mixture distribution comprised of $\alpha$-stable components with mixing weights specified by the chosen frequency distribution. Define the total annual loss random variable as,
\begin{equation}
Z_{t} = Z^{(1)}_{t} + Z^{(2)}_{t} = \sum_{i=1}^{N^{(1)}(t)}X_{i}^{(1)}(t) + \sum_{j=1}^{N^{(2)}(t)}X_{j}^{(2)}(t).
\label{combined}
\end{equation}
Furthermore, if each of these loss processes has an insurance mitigation applied under an ILP policy with top cover limits, $\delta^{(1)} \geq TCL^{(1)}$ and $\delta^{(2)} \geq TCL^{(2)}$, then the closed-form analytic expression for the annual loss density for $N_t^{(1)} + N_t^{(2)} > 0$ is given by
\begin{equation}
\begin{split}
f_{Z_T}(z) &= \sum_{m=0}^{\infty}\sum_{n=0}^{\infty} Pr(N_t^{(1)} = m)Pr(N_t^{(2)} = n) \left[ \sqrt{\frac{\widetilde{\gamma}_{nm}}{2\pi}}\frac{1}{\left(z-\widetilde{\delta}_{nm}\right)^{3/2}}\exp\left(-\frac{\widetilde{\gamma}_{nm}}{2\left(z-\widetilde{\delta}_{nm}\right)}\right) \right] \times \mathbb{I}\left[z>\widetilde{\delta}_{nm}\right]\\
\widetilde{\gamma}_{nm}^{0.5} &= \sum_{i=1}^n |\gamma^{(1)}_i|^{0.5} + \sum_{j=1}^m |\gamma^{(2)}_j|^{0.5} = n|\gamma^{(1)}|^{0.5} + m|\gamma^{(2)}|^{0.5}, \; \; \; \; \widetilde{\beta}_{nm} = 1 \\
\widetilde{\delta}_{nm} &= \sum_{i=1}^n \delta^{(1)}_i + \sum_{j=1}^m \delta^{(2)}_j + \tan \frac{\pi}{4}\left(\widetilde{\gamma}_{nm} - \sum_{j=1}^n \gamma_j^{(1)} - \sum_{k=1}^m \gamma_k^{(2)}\right) \\
&= n\delta^{(1)} + m\delta^{(2)} + \tan \frac{\pi}{4}\left(\left( n|\gamma^{(1)}|^{0.5} + m|\gamma^{(2)}|^{0.5}\right)^2 - n \gamma^{(1)} - m \gamma^{(2)}\right), 
\end{split}
\label{PoissMix2}
\end{equation}
and $f_{Z_T}(0) = \text{Pr}(N_t^{(1)} + N_t^{(2)}=0)$. The exact form of the insurance mitigated annual loss cumulative distribution function is also expressible in closed-form,
\begin{equation}
\begin{split}
\Pr\left(Z_T < z\right) &= F_{Z_T}(z) = \sum_{m=0}^{\infty}\sum_{n=0}^{\infty}  Pr(N_t^{(1)} = m)Pr(N_t^{(2)} = n) \text{erfc}\left( \sqrt{\frac{\widetilde{\gamma}_{nm}}{2\left(z-\widetilde{\delta}_{nm}\right)}}\right)\times \mathbb{I}\left[z>0\right] \\
&+ Pr(N_t^{(1)} + N_t^{(2)}=0).
\end{split}
\label{PoissMix}
\end{equation}}

\subsection{Example 1: Poisson-$\alpha$-Stable Bivariate Risk Aggregation} 
In the case of two Poisson process LDA models one would obtain the result given by
\begin{equation}
\begin{split}
f_{Z_T}(z) &= \sum_{m=0}^{\infty}\sum_{n=0}^{\infty} \exp(-\lambda^{(1)}-\lambda^{(2)})\frac{\left(\lambda^{(1)}\right)^n\left(\lambda^{(2)}\right)^m}{n!m!} \left[ \sqrt{\frac{\widetilde{\gamma}_{nm}}{2\pi}}\frac{1}{\left(z-\widetilde{\delta}_{nm}\right)^{3/2}}\exp\left(-\frac{\widetilde{\gamma}_{nm}}{2\left(z-\widetilde{\delta}_{nm}\right)}\right) \right] \times \mathbb{I}\left[z>\widetilde{\delta}_{nm}\right]\\
\widetilde{\gamma}_{nm}^{0.5} &= \sum_{i=1}^n |\gamma^{(1)}_i|^{0.5} + \sum_{j=1}^m |\gamma^{(2)}_j|^{0.5} = n|\gamma^{(1)}|^{0.5} + m|\gamma^{(2)}|^{0.5}, \; \; \; \; \widetilde{\beta}_{nm} = 1 \\
\widetilde{\delta}_{nm} &= \sum_{i=1}^n \delta^{(1)}_i + \sum_{j=1}^m \delta^{(2)}_j + \tan \frac{\pi}{4}\left(\widetilde{\gamma}_{nm} - \sum_{j=1}^n \gamma_j^{(1)} - \sum_{k=1}^m \gamma_k^{(2)}\right) \\
&= n\delta^{(1)} + m\delta^{(2)} + \tan \frac{\pi}{4}\left(\left( n|\gamma^{(1)}|^{0.5} + m|\gamma^{(2)}|^{0.5}\right)^2 - n \gamma^{(1)} - m \gamma^{(2)}\right), 
\end{split}
\label{PoissMix2}
\end{equation}
and $f_{Z_T}(0) = \text{Pr}(N_t^{(1)} + N_t^{(2)}=0)=\exp(-\lambda_1-\lambda_2)$. The exact form of the insurance mitigated annual loss cumulative distribution function is also expressible in closed-form,
\begin{equation}
\begin{split}
\Pr\left(Z_T < z\right) &= F_{Z_T}(z) = \sum_{m=0}^{\infty}\sum_{n=0}^{\infty} \exp(-\lambda^{(1)}-\lambda^{(2)})\frac{\left(\lambda^{(1)}\right)^n\left(\lambda^{(2)}\right)^m}{n!m!} \text{erfc}\left( \sqrt{\frac{\widetilde{\gamma}_{nm}}{2\left(z-\widetilde{\delta}_{nm}\right)}}\right)\times \mathbb{I}\left[z>0\right] \\
&+ \exp(-\lambda_1 -\lambda_2).
\end{split}
\label{PoissMix}
\end{equation}

\section{Conclusions}
This paper provided novel closed form analytic results for the several interesting classes of OpRisk LDA models based on the family of doubly stochastic $\alpha$-stable LDA models. Thereby providing models with the ability to capture the heavy tailed loss process typical of OpRisk whilst also providing analytic expressions for the compound process annual loss density and distributions as well as the aggregated compound process annual loss models.

Models for the annual loss process were developed in two scenarios, the first based on a loss process treated as continous in time throughout the year with a stochastic intensity parameter, resulting in an inhomogeneous compound Poisson processes annually. The second scenario considered discretization of the annual loss process into monthly increments with dependent time increments as captured by a binomial process with a stochastic probability of success changing annually. 

The resulting LDA annual loss distribution exact representations derived were then analysed and compared to a standard Monte Carlo estimates obtained after simulating $1,000,000$ annual years. The results demonstrated how accurate the exact expressions were in both low and high frequency scenarios even with a small number of terms in the expansions. In addition, we studied the mean square error of the estated distribution functions in each of the six models developed as a function of the truncation. 

Finally, we demonstrated a novel multiple risk result showing how these resuls could be generalized to the multiple risk process setting to allow one to obtain expressions for the annual loss process of an institutions based on individual loss processes from one of the six doubly stochastic families of models developed.

\section{Appendix 1}
\label{Append1}
Simulation of a univariate random variable from a general $\alpha$-stable model can be achieved based on the following steps.

\begin{enumerate}

\item Sample $W \sim \text{Exp}(1)$

\item Sample $U \sim \text{Uniform}\left[\frac{-\pi}{2},\frac{\pi}{2}\right]$

\item Aply transfromation to obtain sample $\bar{y}$ split into a composite:
\begin{enumerate}

\item $\bar{y} = S_{\alpha,\beta} \frac{ \text{sin}\alpha\left(u + B_{\alpha,\beta}\right)}{\left(\text{cos}u\right)^{\frac{1}{\alpha}}} \left[ \frac{\text{cos}\left(u-\alpha\left(u + B_{\alpha,\beta} \right)\right)}{w}\right]^{\frac{1-\alpha}{\alpha}}$ if $\alpha \neq 1$

\item $\bar{y} = \frac{2}{\pi}\left[\left(\frac{\pi}{2} + \beta u\right)\text{tan}u - \beta \ln \frac{\frac{\pi}{2}w \text{cos}u}{\frac{\pi}{2} + \beta u}\right]$ if $\alpha = 1$

\end{enumerate}
where $S_{\alpha,\beta} = \left(1 + \beta^2 \text{tan}^2 \left(\frac{\pi \alpha}{2}\right)\right)^{\frac{1}{\alpha}}$ and $B_{\alpha,\beta} = \frac{1}{\alpha}\arctan\left(\beta\tan\left(\frac{\pi \alpha}{2}\right)\right)$.
In this setting $\bar{y}$ will be distributed by $\mathcal{S}_{\alpha}\left(\beta,1,0\right)$

\item Apply the transformation to obtain sample $y = \gamma \bar{y} + \delta$ with parameters $\mathcal{S}_{\alpha}\left(\beta,\gamma,\delta\right)$

\end{enumerate}

\section{Appendix 2}
\label{Append2}
\textbf{Proof of Theorem 7} 
Given the LDA structure in which the frequency is distributed generically according to $N(t) \sim f(\cdot)$ and the severity model $X_i(t) \sim S(0.5,1,\gamma,\delta;0)$, the exact expressions for the tail asymptotic of the resulting infinite mixture is given by,
\begin{equation*}
\begin{split}
\text{Pr}\left(Z(t)>z_q\right) &\sim 2x^{-0.5}c_{0.5} \sum_{n=1}^{\infty}Pr(N(t)=n) \widetilde{\gamma}_n^{0.5} = C_{\text{Tail}}\sum_{n=1}^{\infty} W_n 
\end{split}
\end{equation*}
where $z_q$ is an upper tail quantile of the annual loss distribution and it will be assumed for simplicity that $\widetilde{\gamma_n}^{0.5} = \sum_{i=1}^n \gamma_i^{0.5} = n|\gamma|^{0.5}$. These results allows us to determine a unique maximum of $\frac{d}{dn}\log(W_n) = 0$ corresponding to the term $n$ with the maximum contribution to the LDA compound processes tail probability for each model as folows:\newline
\underline{Poisson-$\alpha$-Stable.}\newline
Under this model we have $N(t) \sim Po(\lambda)$ which results in the term $W_n = |\gamma|^{0.5}\exp(-\lambda)\frac{\lambda^n}{(n-1)!}$. Now we approximate this function as being continous in $n$ and we differentiate and equate to zero, $\frac{d}{dn}\log(W_n) = 0
$, to find the mode of the terms in the series being summed. We first apply Stirling's approximation to $n! = \Gamma(n+1) \approx \sqrt{2\pi}n^{n+0.5}e^{-n}$ to give 
\begin{equation*}
\begin{split}
\log W_n &= 0.5\log(|\gamma|) - \lambda + n\log(\lambda) - \log((n-1)!)\\
& = 0.5\log(|\gamma|) - \lambda + n\log(\lambda) - \log(\Gamma(n))\\
& \approx 0.5\log(|\gamma|) - \lambda + n\log(\lambda) - (0.5\log(2\pi)+(n+0.5)\log(n)-n)\\
\Rightarrow \; \; \; \frac{d}{dn}\left[\log W_n\right] &= \log(\lambda) - \log(n) - \frac{1}{2n}\\
\Rightarrow \; \; \text{$n$ large} \; \; \frac{d}{dn}\left[\log W_n\right] &\approx  \log(\lambda) - \log(n).\\
\end{split}
\end{equation*}
Equating this to zero and solving for $n$, tells us that if $\lambda \geq 1$, then the maximum occurs at approx $n = \lambda$. If $\lambda < 1$, then max occurs at approx $n=1$. \qed 

\underline{Doubly Stochastic Poisson-Gamma-$\alpha$-Stable.}\newline
Under this model we have $N(t) \sim Po(\lambda)$ and $\lambda \sim \Gamma\left(\alpha,\beta\right)$ which results in the term $W_n = |\gamma|^{0.5}
\frac{(\alpha + n - 1)!}{(\alpha-1)!n!} \left(\frac{\beta}{1+\beta}\right)^{\alpha}\left(\frac{1}{1+\beta}\right)^{n}$. Now we approximate this function as being continous in $n$ and we differentiate and equate to zero, $\frac{d}{dn}\log(W_n) = 0
$, to find the mode of the terms in the series being summed. Again we apply Stirling's approximation $\Gamma(n+1) \approx \sqrt{2\pi}n^{n+0.5}e^{-n}$ to give
\begin{equation*}
\begin{split}
\log W_n &= 0.5\log(|\gamma|) + \log\left(\Gamma(\alpha + n)\right) - \log\left(\Gamma(\alpha)\right) - \log\left(\Gamma(n+1)\right) + \alpha\log(\beta) - (\alpha + n)\log(\beta + 1)\\
&\approx 0.5\log(|\gamma|) + 0.5\log(2\pi) + (n + \alpha - 0.5)\log(n + \alpha - 1) - \alpha + 1 - \log\left(\Gamma(\alpha)\right) - 0.5\log(2\pi) \\
&- (n+0.5)\log(n) + \alpha\log(\beta) - (\alpha + n)\log(\beta + 1)\\
\Rightarrow \; \; \; \frac{d}{dn}\left[\log W_n\right] &= \log(n + \alpha - 1) + \frac{n}{(n + \alpha - 1)} + \frac{(\alpha - 0.5)}{(n + \alpha - 1)} -\log(n) - 1 - \frac{1}(2n)  - \log(\beta + 1)\\
\end{split}
\end{equation*}
Equating this to zero and solving for $n$, ie. solving for the roots $ \log(2) - \log(n) - \frac{1}{2n} = 0$ can be done by an explicit search, which produces $n=2$. \qed 

\underline{negative binomial-$\alpha$-Stable.}\newline
Under this model we have $N(t) \sim NB(r,p)$ which results in the term \\$W_n = |\gamma|^{0.5} \frac{(n+r-1)!}{(n+r-1-n)!(n)!}\left(1-p\right)^r \left(p\right)^{n}$. Now we approximate this function as being continous in $n$ and we differentiate and equate to zero, $\frac{d}{dn}\log(W_n) = 0$, to find the mode of the terms in the series being summed. We first apply Stirling's approximation to $n! = \Gamma(n+1) \approx \sqrt{2\pi}n^{n+0.5}e^{-n}$ to give 
\begin{equation*}
\begin{split}
\log W_n &= 0.5\log(|\gamma|) + \log((n+r-1)!)-\log((r-1)!) - \log((n)!) + r\log(1-p) + n\log(p)\\
&\approx 0.5\log(|\gamma|) + 0.5\log(2\pi) + (n+r-0.5)\log(n+r-1) -(n+r-1) - \log((r-1)!)\\ 
&- 0.5\log(2\pi) - (n+0.5)\log(n) + n + r\log(1-p) + n\log(p)\\
\Rightarrow \; \; \; \frac{d}{dn}\left[\log W_n\right] &= \log(n+r-1) + \frac{n}{n+r-0.5} +   \frac{r-0.5}{n+r-1} - 1 - \log(n) - \frac{1}{2n} + \log(p)\\
\end{split}
\end{equation*}
Then for a given set of values $r$ and $p$ in the model we can equate to zero this expresion and solve for $n$. \qed 

\underline{Doubly Stochastic negative binomial- Beta-$\alpha$-Stable.}\newline
Under this model we have $N(t) \sim NB\left(r,p\right)$ and $p \sim Be\left(\alpha,\beta\right)$ which results in the term $W_n = |\gamma|^{0.5}\frac{(n+r-1)!p^{(\alpha + r -1)}B(\alpha + rn, \beta + n)}{(r-1)!n!p^{(\alpha + rn -1)}B(\alpha,\beta)}$. 
We will now utilise the well known identity for the Beta function with respect to Gamma functions and a Stirling Approximation given by
$$B(x,y) \approx \sqrt{2 \pi} \frac{x^{x-0.5}y^{y-0.5}}{(x+y)^{x+y-0.5}}$$
Now we approximate this function as being continous in $n$ and we differentiate and equate to zero, $\frac{d}{dn}\log(W_n) = 0$, to find the mode of the terms in the series being summed. 
\begin{equation*}
\begin{split}
\log W_n & = 0.5\log(|\gamma|) + \log(\Gamma(n+r)) + (\alpha + r -1)\log(p) + \log(B(\alpha + rn, \beta + n)) \\
&- \log(\Gamma(r)) - \log(\Gamma(n+1)) - (\alpha + rn -1)\log(p) - \log(B(\alpha,\beta))\\
& \approx 0.5\log(|\gamma|) + 0.5\log(2\pi) + (n+r+0.5)\log(n+r) - n-r + (\alpha + r -1)\log(p) \\
&+ 0.5\log(2\pi) + (\alpha + rn-0.5)\log(\alpha + rn) + (\beta + n-0.5)\log(\beta + n) \\
&- (\alpha + rn + \beta + n -0.5)\log(\alpha + rn+\beta + n) - \log(\Gamma(r)) - 0.5\log(2\pi)\\ 
&- (n+1+0.5)\log(n+1) +n+1 - (\alpha + rn -1)\log(p) - \log(B(\alpha,\beta))\\
\Rightarrow \; \; \; \frac{d}{dn}\left[\log W_n\right] &=  \log(n+r) + \frac{(n+r+0.5)}{(n+r)} + r\log(\alpha + rn) + \frac{r(\alpha + rn-0.5)}{(\alpha + rn)} + \log(\beta + n)\\ 
&+ \frac{(\beta + n-0.5)}{(\beta + n)} - (r+1)\log(\alpha + rn+\beta + n) - \frac{(r+1)(\alpha + rn + \beta + n -0.5)}{(\alpha + rn+\beta + n)}\\ 
&- \log(n+1) - \frac{(n+1+0.5)}{(n+1)} - r\log(p)\\
\end{split}
\end{equation*}
Then for a given set of values $r$ and $p$ in the model we can equate to zero this expresion and solve for $n$. \qed 

\end{document}